%% file: main.tex
\documentclass[11pt]{article}
\input{macros}

\title{Persuasion by Parts}
\author{%
  \authorblock{Shuyuan Jia}{Shanghai University of Finance and Economics}{shuyuan.j@outlook.com}%
  \hspace{2em}%
  \authorblock{Kangning Wang}{Rutgers University}{kn.w@rutgers.edu}%
  \hspace{2em}%
  \authorblock{Renzhe Xu}{Shanghai University of Finance and Economics}{xurenzhe@sufe.edu.cn}%
}
\date{}

\begin{document}
\paperfrontmatter{%
\begin{abstract}
Classical Bayesian persuasion assumes that an information designer can condition signals on the entire state. We propose \emph{persuasion by parts}. In this model, the receiver's payoff from acting is \(X = \sum_{i=1}^n X_i\) for independent coordinates \(X_1,\ldots,X_n\) observed by separate sources. The designer commits in advance to a product experiment, in which each source \(i\) can condition its signal only on \(X_i\). Given the signals, the receiver computes her posterior mean \(y\) of \(X\) and acts with probability \(g(y)\), where \(g\colon \R \to [0,1]\) is a nondecreasing function. The designer aims to maximize the probability of action.

Let \(\Vprod(g)\) and \(\Vjoint(g)\) be the supremum probabilities of action under persuasion by parts and under joint persuasion, respectively. When the designer knows \(g\), we show
\[
  \Vprod(g) \;\ge\; c\,\Vjoint(g)^2
  \qquad\text{for every nondecreasing } g\colon \R \to [0,1],
\]
where \(c > 0\) is an absolute constant independent of \(n\) and the prior; the exponent \(2\) is tight. When \(g\) is unknown, we construct a product experiment \(\Pi\) from the prior alone, whose probability of action \(V_\Pi(g)\) satisfies
\[
  V_\Pi(g) \;\ge\; c_\varepsilon\,\Vjoint(g)^{2 + \varepsilon}
  \qquad\text{for every } \varepsilon > 0 \text{ and every nondecreasing } g\colon \R \to [0,1],
\]
where \(c_\varepsilon > 0\) depends only on \(\varepsilon\).
\end{abstract}%
}

\section{Introduction}

In the classical Bayesian persuasion model of \citelink{Emir Kamenica and Matthew Gentzkow}{KamenicaGentzkow}~\cite{KamenicaGentzkow}, a designer commits in advance to an experiment that observes the state and sends a signal to a receiver, who updates her belief and decides whether to act. The designer persuades the receiver by \emph{pooling}: grouping states in which the receiver would prefer not to act together with states in which she would, in proportions that leave her willing to act. When the state has multiple coordinates, an optimal experiment typically pools across all of them jointly. For instance, a credit evaluation that sees both an applicant's liquidity and collateral can pool applicants who are weak in one coordinate and strong in the other with applicants who are strong in both, using a high value in one coordinate to make up for a low value in the other.

In many settings, however, no single observer sees the entire state. Instead, the quantities that matter for a decision are observed by separate sources that do not see one another's data, while the receiver cares about their sum.

Consider, for example, a buyer deciding whether to adopt a system assembled from components made by \(n\) different suppliers. Each component contributes to a shared budget (such as total weight, power draw, or latency), so the buyer wants to adopt if the sum across components stays within budget. A system vendor wants the buyer to adopt and can specify each supplier's grading rule in advance using the prior distribution of the components, yet each supplier tests only its own component and has no access to the other suppliers' results.

Similarly, suppose a division lead seeks approval from a corporate board for a nationwide launch whose total profit is the sum of returns across \(n\) regional markets. The division lead wants the launch approved, while the board wants to approve only if total profit covers the cost of the launch; each regional branch studies its own market and estimates its return without seeing the others. Likewise, when a sponsor pitches a project to an investor, separate teams may independently evaluate engineering cost, market demand, and operational risk. Whether because of privacy constraints, separate departments, or specialized expertise, no single source sees all the data: the designer can specify the signaling rules in advance using the prior, but each source observes and reports on its own coordinate alone.

We call this setting \emph{persuasion by parts}. Because no source knows whether the other coordinates are high or low when it sends its signal, the designer can no longer use a high value in one coordinate to make up for a low value in another on a state-by-state basis. This leads to our central question:
\begin{quote}
\emph{When the receiver's payoff is a sum of independent coordinates observed by separate sources, how much does the designer lose under persuasion by parts?}
\end{quote}

\subsection{Model and baseline bounds}

Let \(n \ge 1\), write \([n] := \{1,\ldots,n\}\), and let \(X_1,\ldots,X_n\) be independent real random variables with finite means and distributions \(P_1,\ldots,P_n\). Set
\[
  P \;:=\; \bigotimes_{i=1}^n P_i
  \qquad\text{and}\qquad
  X \;:=\; \sum_{i=1}^n X_i.
\]
We call \(P\) the \emph{product prior}, \(X_i\) \emph{coordinate \(i\)}, and its observer \emph{source \(i\)}. The receiver's payoff from acting is \(X\). Throughout, we assume that all sets and functions are measurable; we use \(\Pp\) and \(\E\) for probability and expectation over the state and all signaling randomness.

After seeing all signals, the receiver computes her posterior mean \(y\) of \(X\) and acts with probability \(g(y)\), where the \emph{decision function} \(g\colon \R \to [0,1]\) is nondecreasing, and the designer aims to maximize the probability of action. The basic special case is the family of \emph{cutoff rules} \(y \mapsto \1_{\{y \ge t\}}\) for \(t \in \R\): a receiver with cutoff \(t\) acts if and only if her posterior mean of \(X\) is at least \(t\). More generally, \(g\) can model a receiver whose private cutoff is independent of the state and has distribution function \(g\)~\cite{RayoSegal,KolotilinMylovanovZapechelnyukLi}, a population of receivers with heterogeneous cutoffs, or any nondecreasing payoff to the designer (normalized to \([0,1]\)) that depends on the posterior mean~\cite{GentzkowKamenicaRS,DworczakMartini}.

\begin{definition}[Joint and product experiments]
A \emph{joint experiment} \(\Pi(\mathrm{d}s \mid x_1,\ldots,x_n)\) maps each state in \(\R^n\) to a distribution over signals; its signal \(S\) induces the posterior mean \(Y_\Pi := \E[X \mid S]\). A \emph{product experiment} \(\Pi = (\pi_1,\ldots,\pi_n)\) consists of \emph{signaling rules} \(\pi_i(\mathrm{d}s_i \mid x_i)\) that are run with independent private randomness:
\[
  \Pi(\mathrm{d}s_1,\ldots,\mathrm{d}s_n \mid x_1,\ldots,x_n)
  \;=\; \bigotimes_{i=1}^n \pi_i(\mathrm{d}s_i \mid x_i).
\]
\end{definition}

The designer chooses the signaling rules \(\pi_1,\ldots,\pi_n\) with full knowledge of \(P\); the only restriction is that when the experiment runs, source \(i\) observes only \(X_i\) and its own private randomness. Under a product experiment, the pairs \((X_1,S_1),\ldots,(X_n,S_n)\) are independent, so the posterior mean is additive:
\[
  Y_\Pi \;=\; \E[X \mid S_1,\ldots,S_n] \;=\; \sum_{i=1}^n \E[X_i \mid S_i].
\]
We may therefore assume without loss of generality that source \(i\) directly reports its posterior mean \(Y_i := \E[X_i \mid S_i]\) (a \emph{canonical signal}~\cite{GentzkowKamenicaRS}), and that the receiver simply adds up these posterior means. In the language of mean-preserving contractions (equivalently, convex order or majorization~\cite{Strassen,GentzkowKamenicaRS,KleinerMoldovanuStrack}), the posterior mean under a joint experiment can be any mean-preserving contraction of the sum \(X\), whereas under a product experiment it must be a sum of independent mean-preserving contractions of the coordinates \(X_1,\ldots,X_n\). A joint experiment adds, then contracts; a product experiment contracts, then adds.

For a decision function \(g\), we define the \emph{value} of an experiment \(\Pi\), the \emph{product value}, and the \emph{joint value} by
\[
  V_\Pi(g) \;:=\; \E\bigl[g(Y_\Pi)\bigr],
  \qquad
  \Vprod(g) \;:=\; \sup_{\Pi\text{ product}} V_\Pi(g),
  \qquad
  \Vjoint(g) \;:=\; \sup_{\Pi\text{ joint}} V_\Pi(g).
\]
For a cutoff \(t \in \R\), we write \(V_\Pi(t) := \Pp(Y_\Pi \ge t)\) for the value under the cutoff rule \(y \mapsto \1_{\{y \ge t\}}\), and define \(\Vprod(t)\) and \(\Vjoint(t)\) in the same way; all three are nonincreasing in \(t\). Since every product experiment is also a joint experiment, we always have \(\Vprod(g) \le \Vjoint(g)\).

Two simple product experiments are useful benchmarks: the \emph{uninformative experiment} reveals nothing and induces \(Y_\Pi = \E[X]\), whereas \emph{full revelation} reveals every coordinate and induces \(Y_\Pi = X\). By Jensen's inequality, the uninformative experiment is optimal among joint experiments when \(g\) is concave on an interval containing the support of \(X\), and full revelation is optimal when \(g\) is convex on that interval; in either case, \(\Vprod(g) = \Vjoint(g)\). For cutoff rules, which are neither concave nor convex, \(\Vprod(t) < \Vjoint(t)\) is possible even with two identical binary sources.

\begin{example}[Two binary sources]
Let \(X_1\) and \(X_2\) be independent and uniform on \(\{1, -2\}\), and let \(t = 0\). A joint experiment can pool the three states \((1,1)\), \((1,-2)\), and \((-2,1)\), inducing posterior mean \((2 - 1 - 1)/3 = 0\) with probability \(3/4\), so \(\Vjoint(0) = 3/4\).

Under a product experiment, if both sources pool \(X_i = 1\) (mass \(1/2\)) with half of \(X_i = -2\) (mass \(1/4\)) to report posterior mean \(0\) with probability \(3/4\), the receiver acts only when both sources pool, which occurs with probability \((3/4)^2 = 9/16\). A better (and in fact optimal) product experiment is asymmetric: one source reveals nothing (reporting its prior mean \(Y_1 = -1/2\)), while the other pools \(X_2 = 1\) (mass \(1/2\)) with one-fifth of \(X_2 = -2\) (mass \(1/10\)) to report posterior mean \(Y_2 = 1/2\) with probability \(3/5\) and reveals \(X_2 = -2\) otherwise, achieving \(\Vprod(0) = 3/5 < 3/4 = \Vjoint(0)\).
\end{example}

\paragraph{The joint cutoff value and greedy coordinatewise pooling.}
For a real random variable \(Z\) with finite mean and \(u\)-quantile \(F_Z^{-1}(u) := \inf\bigl\{z \in \R : \Pp(Z \le z) \ge u\bigr\}\), its \emph{top-\(s\) pool} selects the top \(s\) probability mass of \(Z\), namely all values above \(F_Z^{-1}(1 - s)\) together with a fraction at \(F_Z^{-1}(1 - s)\) if needed to bring the total mass to \(s\). We write \(w_s(z) \in [0,1]\) for the resulting \emph{top-\(s\) selector} of \(Z\) and call the mean of the top-\(s\) pool
\[
  m_Z(s) \;:=\; \frac{\E\bigl[w_s(Z)\,Z\bigr]}{s} \;=\; \frac{1}{s}\int_{1-s}^1 F_Z^{-1}(u)\,\mathrm{d}u
\]
the \emph{pool mean}, which is continuous and nonincreasing in \(s \in (0,1]\). Because the top-\(s\) pool selects the largest values of \(Z\), its mean \(m_Z(s)\) is the highest conditional mean achievable on any event of probability \(s\), which gives the following characterization of \(\Vjoint(t)\).

\begin{lemma}[Joint cutoff value]
\label{lem:joint-value}
For every cutoff \(t \in \R\),
\[
  \Vjoint(t) \;=\; \max\bigl\{s \in (0,1] : m_X(s) \ge t\bigr\},
\]
where throughout the maximum over the empty set is \(0\), and the maximum is attained by pooling the top-\(\Vjoint(t)\) mass of \(X\). In particular, for every \(q \in (0,1]\), we have \(\Vjoint(t) \ge q\) if and only if \(m_X(q) \ge t\).
\end{lemma}

We say that source \(i\) \emph{pools its top-\(s\) mass} if it sends a single \emph{pool signal} with probability \(w_s(X_i)\), inducing posterior mean \(m_{X_i}(s)\), and otherwise reveals \(X_i\). A natural product counterpart to the \lemref[joint cutoff value]{lem:joint-value} is to let every source pool its top-\(q\) mass for \(q := \Vjoint(t) > 0\), inducing posterior mean \(\sum_{i=1}^n m_{X_i}(q)\) whenever all \(n\) sources pool. If \(A\) is the event on which \(X\) is selected into its top-\(q\) pool, then \(\Pp(A) = q\) gives \(m_{X_i}(q) \ge \E[X_i \mid A]\) for each \(i\), so \(\sum_{i=1}^n m_{X_i}(q) \ge \E[X \mid A] = m_X(q) \ge t\) and the receiver acts whenever all \(n\) sources pool, which occurs with probability \(q^n\).

\begin{proposition}[Greedy coordinatewise pooling]
\label{prop:greedy}
For every product prior and every \(t \in \R\), \(\Vprod(t) \ge \Vjoint(t)^n\).
\end{proposition}

For \(n = 2\) sources, \propref[greedy coordinatewise pooling]{prop:greedy} gives \(\Vprod(t) \ge q^2\), and this quadratic loss is in fact unavoidable: as we show in the \thmref[two-source quadratic lower bound]{thm:two-source-lower}, already with two sources \(\Vprod(t)\) can be arbitrarily close to \(q^2\) while \(\Vjoint(t) \ge q\). As the number of sources \(n\) grows, however, the greedy bound \(q^n\) vanishes exponentially for any fixed \(q < 1\), leaving open whether \(\Vprod(t)\) is even bounded away from zero independently of \(n\).

\subsection{Our results}

We show that the product value is always at least quadratic in the joint value when \(g\) is known, and almost quadratic when a single product experiment must work for all decision functions \(g\) simultaneously.

\begin{introtheorem}[Tight guarantees for persuasion by parts]
For every product prior \(P = \bigotimes_{i=1}^n P_i\):
\begin{enumerate}[label=(\roman*)]
\item \thmref{thm:known-guarantee}: when the decision function \(g\) is known,
\[
  \Vprod(g) \;\ge\; \frac{1}{144}\,\Vjoint(g)^2
  \qquad\text{for every nondecreasing } g\colon \R \to [0,1].
\]
\item \thmref{thm:universal-guarantee}: when the decision function is unknown, there is a single product experiment \(\Pi\), determined by \(P\) alone, such that for every \(\varepsilon > 0\),
\[
  V_\Pi(g) \;\ge\; c_\varepsilon\,\Vjoint(g)^{2 + \varepsilon}
  \qquad\text{for every nondecreasing } g\colon \R \to [0,1],
\]
where \(c_\varepsilon > 0\) depends only on \(\varepsilon\).
\end{enumerate}
\end{introtheorem}

Part~(i), and specifically the single-cutoff bound \(\Vprod(t) \ge \frac{1}{36}\,\Vjoint(t)^2\) in \cref{sec:known}, is the main result of the paper: it shows that product experiments always achieve at least a constant times the square of the joint value for every number \(n\) of sources and arbitrary distributions \(P_i\) with finite means. Equivalently, for any independent real random variables \(X_1,\ldots,X_n\) with finite means, any mean-preserving contraction \(Y\) of \(\sum_{i=1}^n X_i\), and any \(t \in \R\), there exist independent mean-preserving contractions \(Y_i\) of \(X_i\) with \(\Pp\bigl(\sum_{i=1}^n Y_i \ge\nobreak t\bigr) \ge \frac{1}{36}\,\Pp(Y \ge t)^2\).

In Part~(ii) (\cref{sec:universal}), we show that a single product experiment \(\Pi\) simultaneously achieves almost the same quadratic bound across all decision functions \(g\); in fact, for \(q := \Vjoint(g)\), the \(q^\varepsilon\) loss relative to \(q^2\) can be sharpened to an \(\exp\Bigl(-O\bigl(\ln(1/q)/\ln\ln(1/q)\bigr)\Bigr)\) factor (\thmref[universal guarantee]{thm:universal-guarantee}). Moreover, for explicitly given discrete priors, both guarantees can be achieved in polynomial time (\thmref[polynomial-time construction]{thm:polytime} in \cref{sec:computation}).

Finally, the lower bounds and impossibility results in \cref{sec:limitations} show that the exponent \(2\) is tight in both parts, and that neither independence nor additivity can be dropped in general.
\begin{itemize}
\item \thmref{thm:two-source-lower}: even for two sources and a known cutoff, \(\Vprod(t)\) can be arbitrarily close to \(\Vjoint(t)^2\), so Part~(i) is tight up to the constant factor and having many sources is at most a constant factor worse than having two.
\item \thmref{thm:no-universal-quadratic}: Part~(ii) cannot be improved to \(c\,\Vjoint(g)^2\) with a constant \(c > 0\) independent of \(n\), even across two cutoffs, and the sub-polynomial factor in the \thmref[universal guarantee]{thm:universal-guarantee} is tight up to the constant in the exponent.
\item \propref{prop:discontinuity-at-one}: although \(\Vjoint(t) = 1\) implies \(\Vprod(t) = 1\), an application of \thmref{thm:feige}~\cite{Feige,FuEtAl,NieWei} gives, for every \(\varepsilon > 0\), an instance with \(\Vjoint(t) \ge 1 - \varepsilon\) and \(\Vprod(t) \le 1 - 1/\ee\).
\item \propref{prop:correlated-prior-impossibility}: under correlated priors, \(\Vprod(t)\) can be arbitrarily small even when \(\Vjoint(t) = 1/2\) with two sources.
\item \propref{prop:xos-impossibility}: for a monotone symmetric XOS payoff of \(n\) independent coordinates, \(\Vprod(t)\) can be exponentially small in \(n\) even when \(\Vjoint(t) = 1/2\).
\end{itemize}

\subsection{Technical overview}

The first step is to write any decision function \(g\) as a mixture of cutoff rules, \(g(y) = \int_0^1 \1_{\{g(y) > u\}}\,\mathrm{d}u\), reducing both parts of our main theorem to single cutoffs (\lemref[cutoff reduction for a known decision function]{lem:known-g-reduction} and \lemref[cutoff reduction for a fixed experiment]{lem:fixed-exp-reduction}).

\paragraph{Pooling capacity and sparse pooling (\cref{sec:known}).}
Fix a cutoff \(t\) with joint value \(q := \Vjoint(t) > 0\). Under greedy coordinatewise pooling, every source pools its top-\(q\) mass at once; when \(n\) is large, most sources fall outside their pools and reveal values below their means, lowering the sum. To avoid having multiple sources pool at the same time, our \lemref[single-cutoff experiment]{lem:single-cutoff} uses \emph{sparse pooling}, in which typically at most one source pools while the other \(n - 1\) sources reveal their coordinates. For each source, its \emph{target} is the value needed to reach \(t\) when the sum of the other coordinates equals its \((1 - q/3)\)-quantile, and its \emph{pooling capacity} is the largest mass it can pool with mean at least this target. When the total pooling capacity is at least \(q/6\), we scale the capacities down so that the pooling probabilities sum to \(q/6\) and let each source pool its corresponding top mass. Whenever a source pools, the sum of the other coordinates is at least that quantile with probability at least \(q/3\) and another source also pools with probability at most \(q/6\), so a single source pools and induces action with total probability at least \((q/6)(q/3 - q/6) = q^2/36\).

If instead the total pooling capacity is below \(q/6\), every source reveals its coordinate, and the \lemref[capacity lemma]{lem:capacity} shows that full revelation alone gives \(\Pp(X \ge t) > q^2/36\). Like the core--tail decomposition for an additive buyer with independent item values~\cite{LiYao,HartNisanApproximation,BabaioffImmorlicaLucierWeinberg,CaiDevanurWeinberg}, the proof caps each coordinate at its target and splits the expected surplus of \(X\) above \(t\) into a tail term (from coordinates exceeding their targets) and a core term (from the capped sum). Comparing against the top-\(q\) pool of \(X\) bounds the tail term in terms of the total pooling capacity, while revealing the coordinates one by one until the conditional probability of \(X \ge t\) reaches \(q/3\) bounds the core term in terms of \(\Pp(X \ge t)\), showing that the total pooling capacity and \(\Pp(X \ge t)\) cannot both be small.

\paragraph{Universal persuasion across dyadic levels (\cref{sec:universal}).}
When \(g\) is unknown, the experiment cannot be tailored to a single cutoff \(t\). Instead, we consider the cutoffs \(t_j := m_X(2^{-j})\) for \(j \ge 1\), which cover every cutoff up to a factor of \(2\) in the joint value. Our \defref[universal experiment]{def:universal-experiment} mixes over the single-cutoff experiments for all dyadic levels: each source privately draws a level \(j \ge 1\) with probability \(\frac{1}{2j(j+1)}\) and runs that level's signaling rule, or reveals its coordinate directly with probability \(1/2\). Because each source can pool with probability up to \(1/2\) at a given value, the probability that all \(n\) sources reveal at a given state can be as small as \(2^{-n}\), even though the expected number of pooling sources is at most \(1/24\) across all levels. Using a moment bound for products of independent factors (\lemref[conditional product bound]{lem:conditional-product-bound}), we show that mixing over all levels loses at most a sub-polynomial factor.

\subsection{Related work}

\paragraph{Posterior-mean and multi-source persuasion.}
A large literature on posterior-mean persuasion~\cite{KamenicaGentzkow,GentzkowKamenicaRS} characterizes optimal experiments using linear-programming duality and the extreme points of the set of mean-preserving contractions of a single distribution~\cite{Kolotilin,DworczakMartini,KleinerMoldovanuStrack,ArieliBabichenkoSmorodinskyYamashita,DworczakKolotilin}. In our setting, the product constraint instead requires a sum of \emph{independent} contractions. Previous work on independent signaling across independent coordinates has focused on unit-demand settings, where the receiver picks the best of \(n\) independent options rather than adding them up: \citelink{Shaddin Dughmi and Haifeng Xu}{DughmiXu}~\cite{DughmiXu} give a \((1 - 1/\ee)\)-approximation when the \(n\) options are i.i.d., and \citelink{Niloufar Mirzavand Boroujeni, Krishnamurthy Iyer, and William L.~Cooper}{MirzavandIyerCooper}~\cite{MirzavandIyerCooper} prove a tight \(1 - (1 - 1/n)^n \ge 1 - 1/\ee\) guarantee for persuading a receiver to join one of \(n\) independent service locations. When the receiver acts once \(\max_{i \in [n]} Y_i\) (rather than \(\sum_{i=1}^n Y_i\)) reaches a cutoff, as in~\cite{MirzavandIyerCooper}, a low posterior mean on one coordinate is simply ignored, so each source can pool its upper tail without hurting the others. Under a sum, by contrast, every low posterior mean lowers the sum, and even two sources suffer a quadratic loss. Moreover, we study universal experiments that work for all decision functions at once.

\citelink{Itai Arieli, Ronen Gradwohl, and Yishu Zeng}{ArieliGradwohlZeng}~\cite{ArieliGradwohlZeng} study decentralized persuasion by senders who observe conditionally independent signals about a common state that their combined information (almost) fully reveals. For a binary state and enough senders, they get a prior-dependent constant-factor approximation (and full recovery using shared randomness hidden from the receiver). In our setting, by contrast, the sources observe independent coordinates of a sum rather than noisy signals of a common state, no constant-factor approximation is possible even for two sources, and our quadratic guarantee holds uniformly in \(n\) and the prior. Other multi-source persuasion models have studied distributed mediators in auctions~\cite{FeldmanTennenholtzWeinstein} and feasible joint posterior distributions for a binary state~\cite{ArieliBabichenkoSandomirskiyTamuz}.

\paragraph{Approximation in information design.}
On the algorithmic side (see the survey by \citelink{Shaddin Dughmi}{DughmiSurvey}~\cite{DughmiSurvey}), computing an optimal experiment is often computationally hard, even approximately~\cite{DughmiHardness,XuPublicPersuasion,DughmiXu}. This has led to approximation schemes~\cite{ChengEtAl}, constant-factor guarantees for monotone submodular objectives with multiple receivers~\cite{BabichenkoBarman,DughmiXuNoExternalities}, and approximate signaling schemes in auctions~\cite{DughmiImmorlicaRoth,DaskalakisPapadimitriouTzamos,AlijaniBanerjeeMunagalaWang,CaiLiWu}. While our main focus is on how much value is lost under persuasion by parts, our constructions can also be implemented in polynomial time (\cref{sec:computation}).

\paragraph{Universal experiments and receiver uncertainty.}
In a binary-state multi-receiver setting without externalities, a single private experiment can \((1 - 1/\ee)\)-approximate every monotone submodular objective~\cite{DughmiXuNoExternalities}. Approximate majorization~\cite{GoelMeyerson} has also been used for simultaneous optimization in fair price discrimination and fair selection~\cite{BanerjeeMunagalaShenWangFairPrice,BanerjeeMunagalaShenWang}, where the objective depends on the receivers' expected utilities rather than the posterior mean. A related literature studies persuasion under receiver uncertainty through priors over cutoffs~\cite{RayoSegal,KolotilinMylovanovZapechelnyukLi}, worst-case guarantees~\cite{DworczakPavan}, online regret~\cite{BabichenkoTalgamCohenXuZabarnyiRegret,CastiglioniCelliMarchesiGatti}, quantal response~\cite{FengHoTang}, preference perturbations~\cite{GradwohlHuSmorodinsky}, and quota rules~\cite{BergemannGanLi}. Our universal guarantee instead compares a single experiment directly against \(\Vjoint(g)\) for every decision function \(g\).

\paragraph{Strategic senders and constrained channels.}
Finally, our model differs from strategic multi-sender persuasion~\cite{GentzkowKamenicaCompetition,GentzkowKamenicaMultipleSenders,AuKawai,LiNorman,HossainEtAl}, including models in which strategic senders observe separate coordinates or independent options~\cite{DingFengHoTangXu,BanerjeeMunagalaShenWangCompetitive,TejaGhalmeGujar}, in that our sources have no incentive constraints: a single designer commits to all signaling rules. Similarly, whereas prior work on constrained persuasion restricts whether signals are public or private~\cite{BabichenkoBarman,ArieliBabichenko,XuPublicPersuasion} or how much communication each sender can use~\cite{DughmiImmorlicaRoth,DughmiKempeQiang,GradwohlEtAl}, our constraint is simply that each source signals independently based on its own coordinate alone.

\section{Known decision functions}
\label{sec:known}

We begin with the setting where the designer knows the decision function \(g\), and show that product experiments achieve a quadratic guarantee for any number of sources and any product prior.

\begin{theorem}[Quadratic guarantee for a known decision function]
\label{thm:known-guarantee}
For every product prior \(P = \bigotimes_{i=1}^n P_i\), every cutoff \(t \in \R\), and every decision function \(g\),
\[
  \Vprod(t) \;\ge\; \frac{1}{36}\,\Vjoint(t)^2
  \qquad\text{and}\qquad
  \Vprod(g) \;\ge\; \frac{1}{144}\,\Vjoint(g)^2.
\]
\end{theorem}

We first reduce general decision functions to cutoff rules, losing at most a factor of \(4\).

\begin{lemma}[Cutoff reduction for a known decision function]
\label{lem:known-g-reduction}
Let \(c > 0\). If \(\Vprod(t) \ge c\,\Vjoint(t)^2\) for every \(t \in \R\), then
\[
  \Vprod(g) \;\ge\; \frac{c}{4}\,\Vjoint(g)^2
  \qquad\text{for every decision function } g.
\]
\end{lemma}

\begin{proof}
Let \(Y\) be the posterior mean induced by a joint experiment. Since \(g(Y_\Pi) \ge g(t)\,\1_{\{Y_\Pi \ge t\}}\) for every product experiment \(\Pi\) and \(t \in \R\), the assumption \(\Vprod(t) \ge c\,\Vjoint(t)^2\) gives
\[
  \Vprod(g) \;\ge\; g(t)\,\Vprod(t) \;\ge\; c\,g(t)\,\Pp(Y \ge t)^2.
\]
Since \(g\) is nondecreasing, taking the supremum over \(t\) with \(g(t) > u\) gives \(\Vprod(g) \ge c\,u\,\Pp\bigl(g(Y) > u\bigr)^2\) for every \(u \in (0,1)\), and integrating over \(u \in (0,1)\) gives
\[
  \E\bigl[g(Y)\bigr]
  \;=\; \int_0^1 \Pp\bigl(g(Y) > u\bigr)\,\mathrm{d}u
  \;\le\; \sqrt{\frac{\Vprod(g)}{c}}\int_0^1 u^{-1/2}\,\mathrm{d}u
  \;=\; 2\sqrt{\frac{\Vprod(g)}{c}}.
\]
Taking the supremum over joint experiments and rearranging completes the proof.
\end{proof}

For the rest of this section, we fix \(q \in (0,1]\) and a cutoff \(t \le m_X(q)\) (so \(\Vjoint(t) \ge q\) by the \lemref[joint cutoff value]{lem:joint-value}), and write
\[
  p \;:=\; \Pp(X \ge t)
\]
for the value of full revelation. Our goal is to construct a product experiment that induces action with probability at least \(q^2/36\).

\subsection{Pooling capacity and sparse pooling}

Suppose that a single source \(i\) pools while the other \(n - 1\) sources reveal their coordinates, so that the receiver's posterior mean of \(X\) is \(Y_i + X_{-i}\), where \(X_{-i} := X - X_i\). For each \(i \in [n]\), we define the \emph{target} \(h_i\), the \emph{pooling capacity} \(b_i\), and the \emph{total pooling capacity} \(B\) by
\[
  h_i \;:=\; t - F_{X_{-i}}^{-1}\Bigl(1 - \frac{q}{3}\Bigr),
  \qquad
  b_i \;:=\; \max\bigl\{s \in (0,1] : m_{X_i}(s) \ge h_i\bigr\},
  \qquad
  B \;:=\; \sum_{i=1}^n b_i.
\]
Here \(h_i\) is the posterior mean that source \(i\) needs for the sum to reach \(t\) when \(X_{-i}\) equals its \((1 - q/3)\)-quantile, and \(b_i\) is the largest mass that source \(i\) can pool with mean at least \(h_i\). In particular, \(X_{-i} \ge t - h_i\) with probability at least \(q/3\); whenever this holds and the other sources reveal, any posterior mean \(Y_i \ge h_i\) from source \(i\) raises the sum to at least \(t\).

\begin{figure}[!t]
\centering
\begin{tikzpicture}[y=1.18cm,font=\small,
    axis/.style={-{Latex[length=4.5pt,width=3.2pt]},draw=ContentsHeading,line width=0.55pt},
    ptr/.style={-{Latex[length=3.5pt,width=2.5pt]},line width=0.45pt},
    card/.style={rounded corners=3pt,line width=0.45pt,inner sep=4.5pt,font=\scriptsize,align=center}]
  \node[font=\fontsize{9.5}{11}\selectfont\bfseries,text=ContentsHeading] at (3.1,2.90) {(a) Top-\(s_i\) pool of \(X_i\)};
  \node[font=\fontsize{9.5}{11}\selectfont\bfseries,text=ContentsHeading] at (11.3,2.90) {(b) Sparse pooling on \(\{X_{-i} \ge t - h_i\}\)};

  \begin{scope}[shift={(0,0)}]
    \fill[RevealAccent!14] (0,0) -- plot[domain=0:3.75,smooth,samples=50] (\x,{1.55*(\x/1.2)*exp(1-\x/1.2)}) -- (3.75,0) -- cycle;
    \fill[PoolAccent!28] (3.75,0) -- plot[domain=3.75:5.75,smooth,samples=35] (\x,{1.55*(\x/1.2)*exp(1-\x/1.2)}) -- (5.75,0) -- cycle;

    \draw[RevealAccent,line width=1.1pt] plot[domain=0:3.75,smooth,samples=50] (\x,{1.55*(\x/1.2)*exp(1-\x/1.2)});
    \draw[PoolAccent,line width=1.1pt] plot[domain=3.75:5.75,smooth,samples=35] (\x,{1.55*(\x/1.2)*exp(1-\x/1.2)});

    \draw[densely dashed,StatementAccent,line width=0.6pt] (3.75,0) -- (3.75,0.60);
    \draw[ContentsHeading,line width=0.5pt] (3.75,0) -- (3.75,-0.06);
    \node[below,font=\scriptsize,text=StatementAccent] at (3.48,-0.04) {\(F_{X_i}^{-1}(1 - s_i)\)};

    \draw[densely dotted,PoolAccent,line width=0.65pt] (4.70,0) -- (4.70,0.30);
    \fill[PoolAccent,draw=white,line width=0.5pt] (4.70,0) circle (1.9pt);
    \node[below,font=\scriptsize,text=PoolAccent] at (4.70,-0.06) {\(h_i\)};

    \draw[axis] (0,0) -- (6.2,0) node[right,text=black] {\(x_i\)};
    \draw[axis] (0,0) -- (0,2.15) node[above,text=black] {density of \(P_i\)};

    \node[font=\scriptsize,text=RevealAccent] at (1.55,0.52) {reveal \(X_i\) (mass \(1 - s_i\))};
    \node[font=\scriptsize,text=PoolAccent,align=center] at (4.55,1.52)
      {top-\(s_i\) pool (\(s_i = \frac{q}{6}\frac{b_i}{B} \le b_i\)):\\[1pt] mean \(m_{X_i}(s_i) \ge h_i\)};
    \draw[ptr,PoolAccent] (4.45,1.10) to[out=-105,in=70] (4.18,0.34);
  \end{scope}

  \begin{scope}[shift={(8.2,0)}]
    \node[card,draw=PoolAccent!65,fill=PoolAccent!8,minimum width=3.14cm,minimum height=1.41cm] (src1) at (1.03,1.62)
      {\textbf{Source \(i\vphantom{j}\) pools}\\[1.5pt] prob.\ \(s_i = \frac{q}{6}\frac{b_i}{B}\)\\[1.5pt] mean \(m_{X_i}(s_i) \ge h_i\)};

    \node[card,draw=RevealAccent!65,fill=RevealAccent!8,minimum width=3.14cm,minimum height=1.41cm] (src2) at (5.17,1.62)
      {\textbf{Sources \(j \ne i\) reveal}\\[1.5pt] with \(X_{-i} \ge t - h_i\vphantom{\frac{b_i}{B}}\)\\[1.5pt] prob.\ \(\ge q/3 - q/6 = q/6\vphantom{m_{X_i}}\)};

    \node[card,draw=StatementAccent!65,fill=StatementAccent!8,minimum width=7.28cm,minimum height=0.66cm] (sum) at (3.10,0.30)
      {\textbf{Posterior mean:} \(\;Y_\Pi \;=\; m_{X_i}(s_i) + X_{-i} \;\ge\; h_i + (t - h_i) \;=\; t\)};

    \draw[ptr,PoolAccent!85,line width=0.6pt] (src1.south) -- (src1.south |- sum.north);
    \draw[ptr,RevealAccent!85,line width=0.6pt] (src2.south) -- (src2.south |- sum.north);
    \node[circle,fill=white,draw=StatementAccent!55,line width=0.45pt,inner sep=1.2pt,font=\scriptsize\bfseries,text=StatementAccent] at (3.10,1.62) {\raisebox{-0.12ex}{\(+\)}};
  \end{scope}
\end{tikzpicture}
\caption{The \lemref[single-cutoff experiment]{lem:single-cutoff} when \(B \ge q/6\). (a)~Source \(i\) pools its top-\(s_i\) mass at mean \(m_{X_i}(s_i) \ge h_i\) and reveals all other values. (b)~When only source \(i\) pools and \(X_{-i} \ge t - h_i\), the posterior means sum to at least \(t\).}
\label{fig:sparse-pooling}
\end{figure}
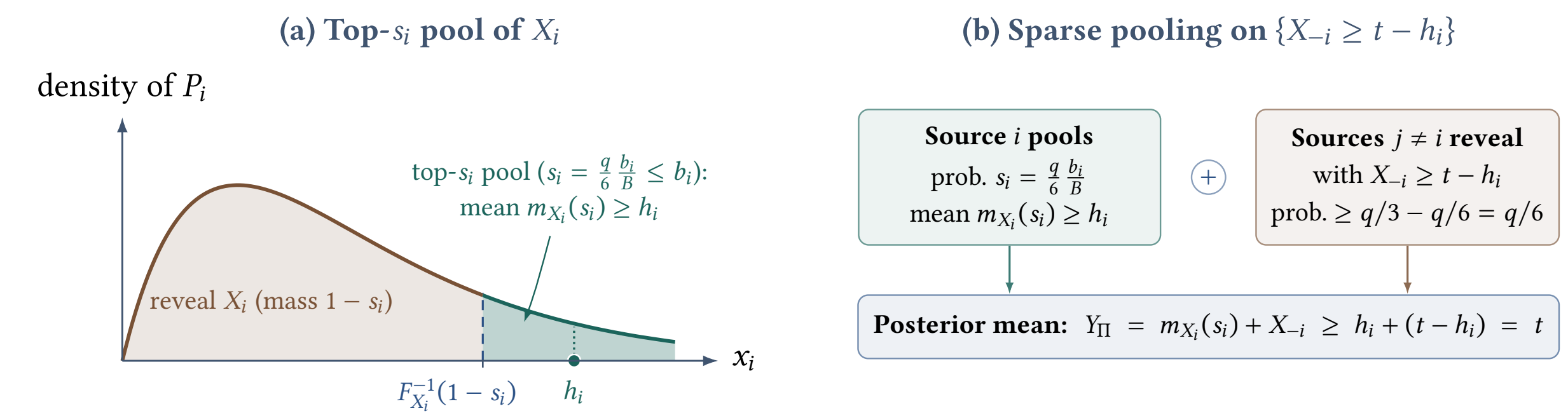

When \(B \ge q/6\), our single-cutoff experiment (\cref{fig:sparse-pooling}) scales the capacities down so that the pooling probabilities sum to \(q/6\) and the expected number of pooling sources is only \(q/6\) (\emph{sparse pooling}). When \(B < q/6\), every source simply reveals \(X_i\), and the \lemref[capacity lemma]{lem:capacity} shows that full revelation alone gives \(p = \Pp(X \ge t) > q^2/36\).

\begin{lemma}[Capacity lemma]
\label{lem:capacity}
The total pooling capacity \(B\) satisfies
\[
  B + \frac{18p}{q} \;\ge\; \frac{2q}{3}.
\]
In particular, if \(B < q/6\), then \(p > \frac{1}{36}\,q^2\).
\end{lemma}

We prove the \lemref[capacity lemma]{lem:capacity} in \cref{sec:proof-capacity}. First, we use it to analyze our single-cutoff experiment.

\begin{lemma}[Single-cutoff experiment]
\label{lem:single-cutoff}
For every \(q \in (0,1]\) and cutoff \(t \le m_X(q)\), let \(\Pi\) be the product experiment in which each source \(i\) pools its top-\(s_i\) mass with \(s_i := \frac{q}{6}\frac{b_i}{B}\) if \(B \ge q/6\), and reveals \(X_i\) if \(B < q/6\). Then the number \(N\) of pooling sources satisfies
\[
  \E[N] \;\le\; \frac{q}{6}
  \qquad\text{and}\qquad
  \Pp\bigl(Y_\Pi \ge t,\ N \le 1\bigr) \;\ge\; \frac{q^2}{36}.
\]
\end{lemma}

\begin{proof}
If \(B < q/6\), then \(\Pi\) is full revelation (\(N = 0\) and \(Y_\Pi = X\)), and the \lemref[capacity lemma]{lem:capacity} gives \(\Pp\bigl(Y_\Pi \ge t,\ N = 0\bigr) = p > \frac{q^2}{36}\).

Now suppose \(B \ge q/6\). Source \(i\) pools with probability \(s_i \le b_i\), so \(\E[N] = \sum_{i=1}^n s_i = q/6\). For each \(i \in [n]\) with \(s_i > 0\), the top-\(s_i\) pool of \(X_i\) has mean \(m_{X_i}(s_i) \ge m_{X_i}(b_i) \ge h_i\); thus, if source \(i\) pools while every other source reveals and \(X_{-i} \ge t - h_i\), then \(N = 1\) and \(Y_\Pi = m_{X_i}(s_i) + X_{-i} \ge t\). Since source \(i\) signals independently of the other sources, applying the union bound over the pooling of sources \(j \ne i\) and summing over the disjoint events across \(i\) gives
\[
  \Pp\bigl(Y_\Pi \ge t,\ N = 1\bigr)
  \;\ge\; \sum_{i:\, s_i > 0} s_i\biggl(\Pp(X_{-i} \ge t - h_i) - \sum_{j \ne i} s_j\biggr)
  \;\ge\; \sum_{i=1}^n s_i\Bigl(\frac{q}{3} - \frac{q}{6}\Bigr)
  \;=\; \frac{q^2}{36}. \qedhere
\]
\end{proof}

Applying the \lemref[single-cutoff experiment]{lem:single-cutoff} with \(q = \Vjoint(t)\) and the \lemref[cutoff reduction for a known decision function]{lem:known-g-reduction} immediately gives the \thmref[quadratic guarantee for a known decision function]{thm:known-guarantee}.

\subsection{Proof of the capacity lemma}
\label{sec:proof-capacity}

\begin{figure}[!t]
\centering
\begin{tikzpicture}[y=1.18cm,font=\small,
    axis/.style={-{Latex[length=4.5pt,width=3.2pt]},draw=ContentsHeading,line width=0.55pt},
    ptr/.style={-{Latex[length=3.5pt,width=2.5pt]},line width=0.45pt},
    dptr/.style={{Latex[length=3.2pt,width=2.3pt]}-{Latex[length=3.2pt,width=2.3pt]},line width=0.45pt},
    card/.style={rounded corners=3pt,line width=0.45pt,inner sep=4pt,font=\scriptsize,align=center}]
  \node[font=\fontsize{9.5}{11}\selectfont\bfseries,text=ContentsHeading] at (3.1,2.90) {(a) Step 2: Balance on the top-\(b_i\) pool};
  \node[font=\fontsize{9.5}{11}\selectfont\bfseries,text=ContentsHeading] at (11.3,2.90) {(b) Step 3: Stopping at \(\tau = i\)};

  \begin{scope}[shift={(0,0)}]
    \fill[RevealAccent!20]
      (3.48,1.30) -- plot[domain=3.48:4.70,smooth,samples=25] (\x,{0.20 + 0.1489*\x + 0.003854*\x*\x*\x}) -- (4.70,1.30) -- cycle;
    \fill[PoolAccent!28]
      (4.70,1.30) -- plot[domain=4.70:5.80,smooth,samples=25] (\x,{0.20 + 0.1489*\x + 0.003854*\x*\x*\x}) -- (5.80,1.30) -- cycle;

    \draw[densely dashed,StatementAccent,line width=0.6pt] (0,1.30) -- (5.80,1.30);
    \node[left,font=\scriptsize,text=StatementAccent] at (0,1.30) {\(h_i\)};

    \draw[densely dashed,ContentsHeading!75,line width=0.55pt] (3.48,0) -- (3.48,1.30);
    \draw[densely dotted,ContentsHeading!75,line width=0.55pt] (5.80,0) -- (5.80,1.816);
    \draw[ContentsHeading,line width=0.5pt] (3.48,0) -- (3.48,-0.06)
      node[below,font=\scriptsize,text=ContentsHeading] {\(1 - b_i\)};
    \draw[ContentsHeading,line width=0.5pt] (5.80,0) -- (5.80,-0.06)
      node[below,font=\scriptsize,text=ContentsHeading] {\(1\)};

    \draw[RevealAccent,line width=1.1pt] plot[domain=0:3.48,smooth,samples=35] (\x,{0.20 + 0.1489*\x + 0.003854*\x*\x*\x});
    \draw[PoolAccent,line width=1.1pt] plot[domain=3.48:5.80,smooth,samples=35] (\x,{0.20 + 0.1489*\x + 0.003854*\x*\x*\x});

    \draw[axis] (0,0) -- (6.2,0) node[right,text=black] {\(u\)};
    \draw[axis] (0,0) -- (0,2.15) node[above,text=black] {\(F_{X_i}^{-1}(u)\)};

    \draw[dptr,RevealAccent!85] (0.65,0.31) -- (0.65,1.29)
      node[midway,right=2.5pt,font=\scriptsize,text=RevealAccent] {larger \((h_i - X_i)_+\)};

    \node[font=\scriptsize,text=RevealAccent,align=center] at (2.35,1.72) {deficit in pool:\\[-1pt] area \(U_i\)};
    \draw[ptr,RevealAccent] (3.08,1.60) to[out=-15,in=135] (3.95,1.16);

    \node[font=\scriptsize,text=PoolAccent,align=center] at (4.35,2.02) {surplus \((X_i - h_i)_+\):\\[-1pt] area \(U_i\)};
    \draw[ptr,PoolAccent] (5.00,1.82) to[out=-25,in=115] (5.48,1.48);
  \end{scope}

  \begin{scope}[shift={(8.2,0)}]
    \draw[densely dashed,StatementAccent,line width=0.6pt] (0,0.85) -- (5.75,0.85);
    \node[left,font=\scriptsize,text=StatementAccent] at (0,0.85) {\(q/3\)};

    \foreach \x/\lbl in {0.50/1, 1.50/2, 3.50/{i-1}, 4.50/{i=\tau}, 5.50/{i+1}} {
      \draw[ContentsHeading,line width=0.5pt] (\x,0) -- (\x,-0.06);
      \node[below,font=\scriptsize,text=ContentsHeading] at (\x,-0.06) {\(\lbl\)};
    }
    \node[below,font=\scriptsize,text=ContentsHeading] at (2.50,-0.06) {\(\vphantom{1}\cdots\)};

    \draw[RevealAccent,line width=1.1pt] (0.50,0.24) -- (1.50,0.52);
    \draw[RevealAccent,densely dotted,line width=1.0pt] (1.50,0.52) .. controls (2.12,0.68) and (2.85,0.18) .. (3.50,0.38);
    \draw[PoolAccent,line width=1.1pt] (3.50,0.38) -- (4.50,1.20);
    \draw[ContentsHeading!55,densely dotted,line width=0.9pt] (4.50,1.20) -- (5.50,1.34);

    \foreach \x/\y in {0.50/0.24, 1.50/0.52, 3.50/0.38} {
      \fill[RevealAccent,draw=white,line width=0.55pt] (\x,\y) circle (2.2pt);
    }
    \fill[PoolAccent,draw=white,line width=0.6pt] (4.50,1.20) circle (2.5pt);
    \fill[ContentsHeading!55,draw=white,line width=0.5pt] (5.50,1.34) circle (2.0pt);

    \draw[axis] (0,0) -- (6.05,0) node[right,text=black] {step \(k\)};
    \draw[axis] (0,0) -- (0,2.15) node[above,text=black] {\(\Pp(X \ge t \mid X_1,\ldots,X_k)\)};

    \node[card,draw=PoolAccent!65,fill=PoolAccent!8] (stopbox) at (3.50,1.82)
      {\textbf{Stop at \(\tau = i\):} \(\;\E\bigl[(C - t)_+ \mid X_1,\ldots,X_i\bigr] \le \phi_i(X_{\le i-1} + h_i)\)};
    \draw[ptr,PoolAccent] ([xshift=1.00cm]stopbox.south) -- (4.50,1.30);

    \node[font=\scriptsize,text=RevealAccent,anchor=west] at (0.18,1.15) {before \(\tau\): \(\;\phi_i(X_{\le i-1} + h_i) \le 9U/q\)};
    \draw[ptr,RevealAccent] (2.45,0.98) to[out=-30,in=135] (3.42,0.45);
  \end{scope}
\end{tikzpicture}
\caption{Proof of the \lemref[capacity lemma]{lem:capacity}. (a)~On the top-\(b_i\) pool (\(u \in [1 - b_i, 1]\), with mean \(h_i = m_{X_i}(b_i)\)), the surplus above \(h_i\) and the deficit below \(h_i\) both have area \(U_i\), while the deficit \((h_i - X_i)_+\) is larger outside the pool. (b)~To bound the core term \(\E\bigl[(C - t)_+\bigr]\) for \(C = \sum_{i=1}^n \min\{X_i, h_i\}\), we reveal \(X_1,\ldots,X_n\) one by one and stop at the first step \(\tau = i\) where \(\Pp(X \ge t \mid X_1,\ldots,X_i) \ge q/3\). On \(E_i = \{\tau = i\}\), \(\min\{X_i, h_i\} \le h_i\) gives \(\E\bigl[(C - t)_+ \mid X_1,\ldots,X_i\bigr] \le \phi_i(X_{\le i-1} + h_i) \le 9U/q\), where \(\phi_i(z) := \E\bigl[(z + X_{>i} - t)_+\bigr]\).}
\label{fig:capacity-proof}
\end{figure}
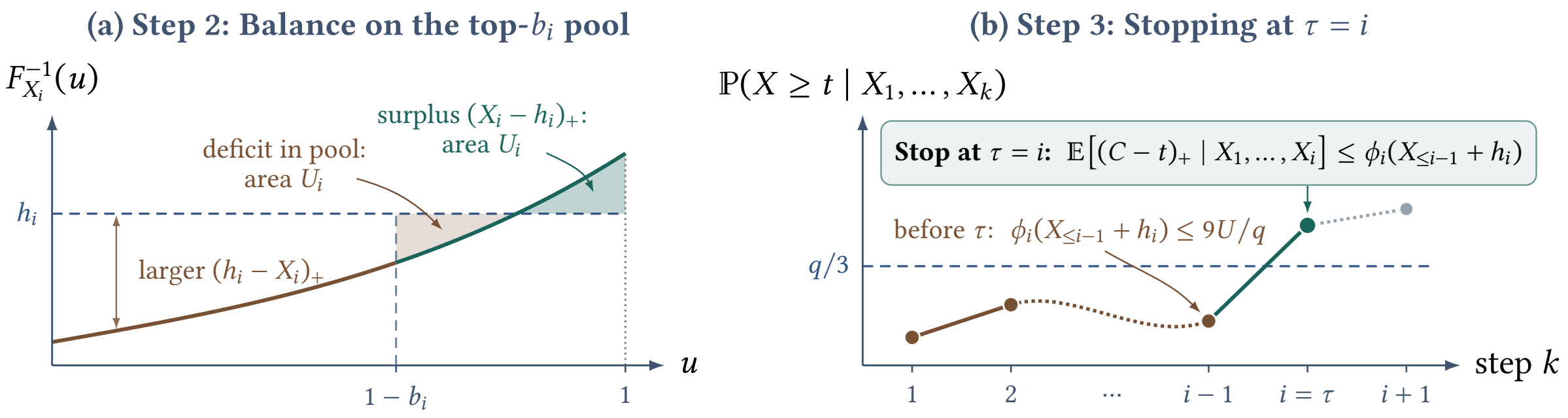

\begin{proof}[Proof of the {\lemref[capacity lemma]{lem:capacity}}]
The bound holds immediately if \(B \ge 2q/3\) or \(p \ge q/3\), so assume \(B < 2q/3\) and \(p < q/3\).

\emph{Step 1: The decomposition.}
Write \(z_+ := \max\{z,0\}\) and let \(U := \E\bigl[(X - t)_+\bigr]\). If \(U = 0\), then \(X \le t\) almost surely, and \(m_X(q) \ge t\) implies that \(X = t\) on the top-\(q\) pool of \(X\). Thus \(p \ge q\), contradicting \(p < q/3\), so \(U > 0\).

Since
\[
  X \;=\; \sum_{i=1}^n \min\{X_i, h_i\} + \sum_{i=1}^n (X_i - h_i)_+,
\]
we have
\begin{equation}
  U \;=\; \E\bigl[(X - t)_+\bigr]
  \;\le\; \underbrace{\sum_{i=1}^n \E\bigl[(X_i - h_i)_+\bigr]}_{\text{tail term}}
  \;+\; \underbrace{\E\Biggl[\biggl(\sum_{i=1}^n \min\{X_i, h_i\} - t\biggr)_+\Biggr]}_{\text{core term}}.
  \label{eq:decomposition}
\end{equation}
We bound the tail term in terms of \(B\) and the core term in terms of \(p\).

\emph{Step 2: Bounding the tail term.}
Let \(U_i := \E\bigl[(X_i - h_i)_+\bigr]\). We show that
\begin{equation}
  \sum_{i=1}^n U_i \;\le\; \frac{3UB}{2q}.
  \label{eq:tail-by-capacity}
\end{equation}

Fix \(i \in [n]\). If \(b_i = 0\), then \(X_i \le h_i\) almost surely and \(U_i = 0\). Otherwise, \(b_i \le B < 2q/3 < 1\) gives \(m_{X_i}(b_i) = h_i\), so the top-\(b_i\) pool has mean \(h_i\) and contains every realization with \(X_i > h_i\). Its surplus and deficit therefore balance (\hyperref[fig:capacity-proof]{\cref*{fig:capacity-proof}a}):
\[
  \frac{U_i}{b_i}
  \;=\; \frac{\E\bigl[w_{b_i}(X_i)(X_i - h_i)_+\bigr]}{b_i}
  \;=\; \frac{\E\bigl[w_{b_i}(X_i)(h_i - X_i)_+\bigr]}{b_i}.
\]

Let \(A\) be the event on which \(X\) is selected into its top-\(q\) pool. Since \(m_X(q) \ge t\),
\begin{equation}
  \E\bigl[\1_A(t - X)_+\bigr]
  \;\le\; \E\bigl[\1_A(X - t)_+\bigr]
  \;\le\; U.
  \label{eq:total-deficit}
\end{equation}
Define \(A_i := A \cap \{X_{-i} \le t - h_i\}\). By the definition of \(h_i\), \(\Pp(X_{-i} > t - h_i) \le q/3\), and on \(A_i\) we have \(h_i - X_i = h_i + X_{-i} - X \le t - X\), so \((h_i - X_i)_+ \le (t - X)_+\). Hence
\[
  \Pp(A_i) \;\ge\; \frac{2q}{3},
  \qquad
  \E\bigl[\1_{A_i}(h_i - X_i)_+\bigr]
  \;\le\; \E\bigl[\1_A(t - X)_+\bigr]
  \;\le\; U.
\]
Since \((h_i - X_i)_+\) is nonincreasing in \(X_i\) and \(\Pp(A_i) \ge 2q/3 > b_i\), the top-\(b_i\) pool of \(X_i\) has a smaller conditional expectation of \((h_i - X_i)_+\) than \(A_i\), so
\[
  \frac{U_i}{b_i}
  \;=\; \frac{\E\bigl[w_{b_i}(X_i)(h_i - X_i)_+\bigr]}{b_i}
  \;\le\; \frac{\E\bigl[\1_{A_i}(h_i - X_i)_+\bigr]}{\Pp(A_i)}
  \;\le\; \frac{3U}{2q}.
\]
Summing over \(i \in [n]\) proves \cref{eq:tail-by-capacity}.

\emph{Step 3: Bounding the core term.}
Let \(C := \sum_{i=1}^n \min\{X_i, h_i\}\). We show that
\[
  \E\bigl[(C - t)_+\bigr] \;\le\; \frac{27Up}{q^2}.
\]
Define the random variable
\[
  \tau \;:=\; \min\bigl\{k \in [n] : \Pp(X \ge t \mid X_1,\ldots,X_k) \ge q/3\bigr\},
\]
where \(\tau = \infty\) if the set is empty, and set \(E_i := \{\tau = i\}\) (\hyperref[fig:capacity-proof]{\cref*{fig:capacity-proof}b}). Since \(C \le X\) and \(X\) is determined by \(X_1,\ldots,X_n\), we have \(\{C \ge t\} \subseteq \{X \ge t\} \subseteq \bigsqcup_{i=1}^n E_i\).

On \(E_i\), the conditional probability of \(X \ge t\) is at least \(q/3\). Since \(E_i\) is determined by \(X_1,\ldots,X_i\),
\[
  p \;=\; \Pp(X \ge t)
  \;=\; \sum_{i=1}^n \Pp\bigl(\{X \ge t\} \cap E_i\bigr)
  \;=\; \sum_{i=1}^n \E\bigl[\1_{E_i}\,\Pp(X \ge t \mid X_1,\ldots,X_i)\bigr]
  \;\ge\; \frac{q}{3}\sum_{i=1}^n \Pp(E_i).
\]
Hence,
\begin{equation}
  \sum_{i=1}^n \Pp(E_i) \;\le\; \frac{3p}{q}.
  \label{eq:stopping-prob}
\end{equation}

For \(i = 0,\ldots,n\), write
\[
  X_{\le i} \;:=\; \sum_{k \le i} X_k,
  \qquad
  X_{>i} \;:=\; \sum_{k > i} X_k,
  \qquad
  \phi_i(z) \;:=\; \E\bigl[(z + X_{>i} - t)_+\bigr].
\]
Each \(\phi_i\) is nonnegative, nondecreasing, and \(1\)-Lipschitz, with \(\E\bigl[\phi_i(X_{\le i})\bigr] = U\).

For every \(i \in [n]\) with \(\Pp(E_i) > 0\), we next show
\begin{equation}
  \E\bigl[(C - t)_+ \mid E_i\bigr] \;\le\; \frac{9U}{q}.
  \label{eq:conditional-core}
\end{equation}
Since
\[
  C
  \;=\; \sum_{k < i}\min\{X_k,h_k\} + \min\{X_i,h_i\} + \sum_{k > i}\min\{X_k,h_k\}
  \;\le\; X_{\le i-1} + h_i + X_{>i}
\]
and \(X_{>i}\) is independent of \(X_1,\ldots,X_i\), we have
\[
  \E\bigl[(C - t)_+ \mid X_1,\ldots,X_i\bigr]
  \;\le\; \phi_i(X_{\le i-1} + h_i).
\]
On \(E_i\), the conditional probability of \(X \ge t\) given \(X_1,\ldots,X_{i-1}\) is below \(q/3\) (for \(i = 1\) by \(p < q/3\), and for \(i > 1\) by the definition of \(\tau\)), so \(z = X_{\le i-1}\) satisfies
\begin{equation}
  \Pp(z + X_{>i-1} \ge t) \;<\; \frac{q}{3}.
  \label{eq:safe-prefix}
\end{equation}
Since \(E_i\) is determined by \(X_1,\ldots,X_i\), to prove \cref{eq:conditional-core} it is enough to show \(\phi_i(z + h_i) \le 9U/q\) for any fixed \(z \in \R\) satisfying \cref{eq:safe-prefix}.

Using \(A\) and \(A_i = A \cap \{X_{-i} \le t - h_i\}\) defined in Step~2, set
\[
  A_{i,z} \;:=\; A_i \cap \{z + X_{>i-1} < t\}.
\]
Since \(\Pp(A_i) \ge 2q/3\), \cref{eq:safe-prefix} gives \(\Pp(A_{i,z}) \ge q/3\). On \(A_{i,z}\),
\[
  h_i - X_i \;\le\; t - X,
  \qquad
  z - X_{\le i-1} \;=\; (z + X_{>i-1}) - X \;<\; t - X.
\]
Adding gives \(z + h_i - X_{\le i} \le 2(t - X)\); since \(\phi_i\) is nondecreasing and \(1\)-Lipschitz,
\[
  \phi_i(z + h_i)
  \;\le\; \phi_i(X_{\le i}) + 2(t - X)_+
  \qquad\text{on } A_{i,z}.
\]
Taking expectations under the prior with \(z\) fixed gives
\[
  \Pp(A_{i,z})\,\phi_i(z + h_i)
  \;\le\; \E\bigl[\1_{A_{i,z}}\,\phi_i(X_{\le i})\bigr]
    + 2\,\E\bigl[\1_{A_{i,z}}(t - X)_+\bigr]
  \;\le\; U + 2U \;=\; 3U.
\]
The last inequality follows from \(A_{i,z} \subseteq A\) and \cref{eq:total-deficit}. Dividing by \(\Pp(A_{i,z}) \ge q/3\) gives \(\phi_i(z + h_i) \le 9U/q\), proving \cref{eq:conditional-core}.

Since \((C - t)_+ = 0\) outside \(\{C \ge t\} \subseteq \bigsqcup_{i=1}^n E_i\), \cref{eq:stopping-prob,eq:conditional-core} give
\[
  \E\bigl[(C - t)_+\bigr]
  \;=\; \sum_{i:\,\Pp(E_i) > 0} \Pp(E_i)\,\E\bigl[(C - t)_+ \mid E_i\bigr]
  \;\le\; \frac{9U}{q}\sum_{i=1}^n \Pp(E_i)
  \;\le\; \frac{9U}{q}\cdot\frac{3p}{q}
  \;=\; \frac{27Up}{q^2}.
\]

\emph{Step 4: Combining the bounds.}
Substituting the tail and core bounds into \cref{eq:decomposition} gives
\[
  U \;=\; \E\bigl[(X - t)_+\bigr]
  \;\le\; \sum_{i=1}^n U_i + \E\bigl[(C - t)_+\bigr]
  \;\le\; \frac{3UB}{2q} + \frac{27Up}{q^2}.
\]
Since \(U > 0\), dividing by \(U\) and rearranging completes the proof.
\end{proof}

\section{Unknown decision functions}
\label{sec:universal}

When the decision function is unknown, a single product experiment must be chosen from the prior alone. In this section, we show that one universal experiment achieves every exponent \(2 + \varepsilon\) simultaneously for all \(\varepsilon > 0\) and all decision functions \(g\).

\begin{theorem}[Universal guarantee]
\label{thm:universal-guarantee}
For every product prior \(P = \bigotimes_{i=1}^n P_i\), there is a product experiment \(\Pi\), determined by \(P\) alone, such that every decision function \(g\) with \(q := \Vjoint(g) > 0\) satisfies
\[
  V_\Pi(g) \;\ge\; c\,q^2 \exp\biggl(-c'\,\frac{\ln(\ee/q)}{\ln\bigl(\ee + \ln(\ee/q)\bigr)}\biggr),
\]
where \(c, c' > 0\) are absolute constants. In particular, for every \(\varepsilon > 0\), there is a constant \(c_\varepsilon > 0\) depending only on \(\varepsilon\) such that \(V_\Pi(g) \ge c_\varepsilon\,\Vjoint(g)^{2 + \varepsilon}\) for every decision function \(g\).
\end{theorem}

As in \cref{sec:known}, we first reduce general decision functions to cutoff rules, but now without losing a constant factor because the same experiment \(\Pi\) works across all cutoffs at once. Indeed, suppose \(V_\Pi(t) \ge c\,\Vjoint(t)^\alpha\) for all \(t \in \R\), where \(c > 0\) and \(\alpha \ge 1\). For any posterior mean \(Y\) induced by a joint experiment, any \(u \in (0,1)\), and any \(t\) with \(g(t) > u\), since \(g\) is nondecreasing we have \(\Pp\bigl(g(Y_\Pi) > u\bigr) \ge V_\Pi(t) \ge c\,\Pp(Y \ge t)^\alpha\). Taking the supremum over \(t\) with \(g(t) > u\) gives \(\Pp\bigl(g(Y_\Pi) > u\bigr) \ge c\,\Pp\bigl(g(Y) > u\bigr)^\alpha\), and integrating over \(u \in (0,1)\) with Jensen's inequality gives \(V_\Pi(g) \ge c\,\E\bigl[g(Y)\bigr]^\alpha\).

\begin{lemma}[Cutoff reduction for a fixed experiment]
\label{lem:fixed-exp-reduction}
Let \(c > 0\) and \(\alpha \ge 1\). If a product experiment \(\Pi\) satisfies \(V_\Pi(t) \ge c\,\Vjoint(t)^\alpha\) for every \(t \in \R\), then \(V_\Pi(g) \ge c\,\Vjoint(g)^\alpha\) for every decision function \(g\).
\end{lemma}

\subsection{The universal experiment}

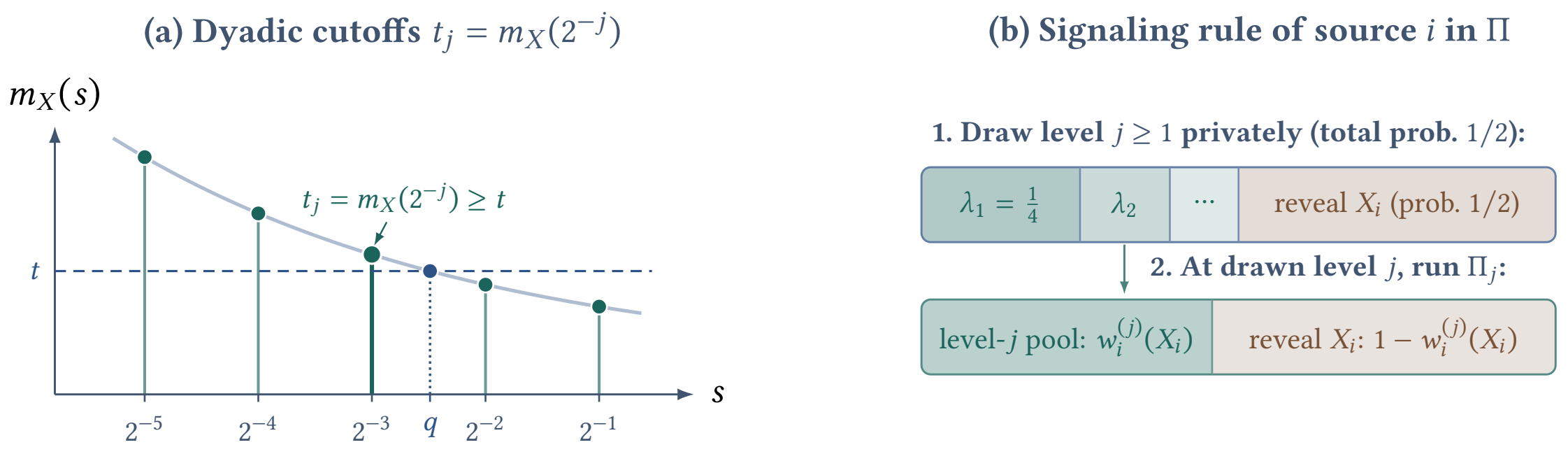
\begin{figure}[!t]
\centering
\begin{tikzpicture}[y=1.18cm,font=\small,
    axis/.style={-{Latex[length=4.5pt,width=3.2pt]},draw=ContentsHeading,line width=0.55pt},
    ptr/.style={-{Latex[length=3.5pt,width=2.5pt]},line width=0.45pt}]
  \node[font=\fontsize{9.5}{11}\selectfont\bfseries,text=ContentsHeading] at (3.1,2.90) {(a) Dyadic cutoffs \(t_j = m_X(2^{-j})\)};
  \node[font=\fontsize{9.5}{11}\selectfont\bfseries,text=ContentsHeading] at (11.3,2.90) {(b) Signaling rule of source \(i\) in \(\Pi\)};

  \begin{scope}[shift={(0,0)}]
    \draw[StatementAccent!38,line width=0.95pt] plot[domain=0.55:5.55,smooth]
      (\x,{0.22 + 2.15*exp(-0.29*\x)});

    \foreach \x/\j/\h in {0.85/5/1.901, 1.925/4/1.450, 3.00/3/1.121, 4.075/2/0.879, 5.15/1/0.703} {
      \draw[ContentsHeading,line width=0.5pt] (\x,0) -- (\x,-0.06);
      \node[below,font=\scriptsize,text=ContentsHeading] at (\x,-0.06) {\(2^{-\j}\)};
      \draw[PoolAccent!65,line width=0.75pt] (\x,0) -- (\x,\h);
      \fill[PoolAccent,draw=white,line width=0.5pt] (\x,\h) circle (2.1pt);
    }
    \draw[PoolAccent,line width=1.15pt] (3.00,0) -- (3.00,1.121);
    \fill[PoolAccent,draw=white,line width=0.6pt] (3.00,1.121) circle (2.5pt);

    \draw[densely dashed,StatementAccent,line width=0.6pt] (0,0.988) -- (5.65,0.988);
    \node[left,font=\scriptsize,text=StatementAccent] at (0,0.988) {\(t\)};
    \draw[densely dotted,StatementAccent,line width=0.65pt] (3.55,0) -- (3.55,0.988);
    \fill[StatementAccent,draw=white,line width=0.5pt] (3.55,0.988) circle (2.1pt);
    \draw[StatementAccent,line width=0.5pt] (3.55,0) -- (3.55,-0.06)
      node[below,font=\scriptsize,text=StatementAccent] {\(q\)};

    \node[font=\scriptsize,text=PoolAccent,align=center] at (3.30,1.56) {\(t_j = m_X(2^{-j}) \ge t\)};
    \draw[ptr,PoolAccent] (3.14,1.38) -- (3.02,1.19);

    \draw[axis] (0,0) -- (6.05,0) node[right,text=black] {\(s\)};
    \draw[axis] (0,0) -- (0,2.15) node[above,text=black] {\(m_X(s)\)};
  \end{scope}

  \begin{scope}[shift={(8.2,0)}]
    \node[above right,font=\scriptsize\bfseries,text=ContentsHeading] at (-0.05,1.86)
      {1.\ Draw level \(j \ge 1\) privately (total prob.\ \(1/2\)):};
    \begin{scope}[shift={(0,1.22)}]
      \clip[rounded corners=2.5pt] (0,0) rectangle (6.0,0.60);
      \fill[PoolAccent!34] (0,0) rectangle (1.50,0.60);
      \fill[PoolAccent!23] (1.50,0) rectangle (2.35,0.60);
      \fill[PoolAccent!14] (2.35,0) rectangle (3.00,0.60);
      \fill[RevealAccent!20] (3.00,0) rectangle (6.0,0.60);
    \end{scope}
    \draw[StatementAccent!75,line width=0.55pt,rounded corners=2.5pt] (0,1.22) rectangle (6.0,1.82);
    \foreach \x in {1.50, 2.35, 3.00} {
      \draw[StatementAccent!65,line width=0.5pt] (\x,1.22) -- (\x,1.82);
    }
    \node[font=\scriptsize,text=PoolAccent] at (0.75,1.52) {\(\lambda_1 = \frac{1}{4}\)};
    \node[font=\scriptsize,text=PoolAccent] at (1.92,1.52) {\(\lambda_2\)};
    \node[font=\scriptsize,text=PoolAccent] at (2.68,1.52) {\(\cdots\)};
    \node[font=\scriptsize,text=RevealAccent] at (4.50,1.52) {reveal \(X_i\) (prob.\ \(1/2\))};

    \draw[ptr,PoolAccent!80,line width=0.55pt] (1.92,1.20) -- (1.92,0.80)
      node[midway,right=3pt,font=\scriptsize\bfseries,text=ContentsHeading] {2.\ At drawn level \(j\), run \(\Pi_j\):};
    \begin{scope}[shift={(0,0.16)}]
      \clip[rounded corners=2.5pt] (0,0) rectangle (6.0,0.60);
      \fill[PoolAccent!30] (0,0) rectangle (2.75,0.60);
      \fill[RevealAccent!16] (2.75,0) rectangle (6.0,0.60);
    \end{scope}
    \draw[PoolAccent!75,line width=0.55pt,rounded corners=2.5pt] (0,0.16) rectangle (6.0,0.76);
    \draw[PoolAccent!65,line width=0.5pt] (2.75,0.16) -- (2.75,0.76);
    \node[font=\scriptsize,text=PoolAccent] at (1.37,0.46) {level-\(j\) pool: \(w_i^{(j)}(X_i)\)};
    \node[font=\scriptsize,text=RevealAccent] at (4.37,0.46) {reveal \(X_i\): \(1 - w_i^{(j)}(X_i)\)};
  \end{scope}
\end{tikzpicture}
\caption{The \defref[universal experiment]{def:universal-experiment}. (a)~Every cutoff \(t\) with \(q = \Vjoint(t) \in (0,1]\) is covered by the dyadic cutoff \(t_j = m_X(2^{-j})\) with \(q/2 \le 2^{-j} \le q\). (b)~Each source \(i\) privately draws a level \(j \ge 1\) with probability \(\lambda_j\) and runs its signaling rule from \(\Pi_j\), or reveals \(X_i\) directly with probability \(1/2\).}
\label{fig:universal-experiment}
\end{figure}

For each integer \(j \ge 1\), define the cutoff \(t_j := m_X(2^{-j})\). For every cutoff \(t \in \R\) with \(q := \Vjoint(t) \in (0,1]\), choosing \(j := \max\Bigl\{1, \bigl\lceil\log_2(1/q)\bigr\rceil\Bigr\}\) gives \(q/2 \le 2^{-j} \le q\), and since \(m_X\) is nonincreasing, the \lemref[joint cutoff value]{lem:joint-value} gives
\begin{equation}
  t_j \;=\; m_X(2^{-j}) \;\ge\; m_X(q) \;\ge\; t.
  \label{eq:dyadic-cutoff}
\end{equation}
To cover all cutoffs at once without shared randomness across sources, our universal experiment (\cref{fig:universal-experiment}) mixes the single-cutoff experiments \(\Pi_j\) for \(t_j\) across all levels \(j \ge 1\) independently at each source.

\begin{definition}[Universal experiment]
\label{def:universal-experiment}
For each \(j \ge 1\), let \(\lambda_j := \frac{1}{2j(j+1)}\), so that \(\sum_{j=1}^\infty \lambda_j = \frac{1}{2}\). Let \(\Pi_j\) be the \lemref[single-cutoff experiment]{lem:single-cutoff} for cutoff \(t_j = m_X(2^{-j})\) (taking \(q = 2^{-j}\)), and write \(w_i^{(j)}(X_i) \in [0,1]\) for the pooling probability of source \(i\) in \(\Pi_j\). In the product experiment \(\Pi\), source \(i\) observes \(X_i\) and, using independent private randomness:
\begin{enumerate}
\item draws a level \(j \in \{1,2,\ldots\}\) with probability \(\lambda_j\), independently of \(X_i\), and otherwise (with probability \(1/2\)) reveals \(X_i\) directly;
\item if it draws level \(j\), runs its signaling rule from \(\Pi_j\): with probability \(w_i^{(j)}(X_i)\), it pools \(X_i\) and sends a level-\(j\) pool signal; otherwise, it reveals \(X_i\).
\end{enumerate}
\end{definition}

Because each source draws its level and pools using its own private randomness, \(\Pi\) is a product experiment, and it is determined by \(P\) alone. Given \(X_i = x_i\), source \(i\) sends its level-\(j\) pool signal with probability \(\lambda_j w_i^{(j)}(x_i)\), so its total pooling probability across all levels is
\[
  \overline{w}_i(x_i) \;:=\; \sum_{j=1}^\infty \lambda_j\,w_i^{(j)}(x_i) \;\le\; \sum_{j=1}^\infty \lambda_j \;=\; \frac{1}{2}.
\]
Because source \(i\) draws its level \(j\) independently of \(X_i\), a level-\(j\) pool signal under \(\Pi\) induces the same posterior mean as the pool signal under \(\Pi_j\), while revealing \(X_i\) induces posterior mean \(X_i\). Since \(\sum_{i=1}^n \E\bigl[w_i^{(j)}(X_i)\bigr] \le \frac{2^{-j}}{6}\) at every level \(j \ge 1\) by the \lemref[single-cutoff experiment]{lem:single-cutoff}, the expected number of pooling sources across all levels satisfies
\begin{equation}
  \sum_{i=1}^n \E\bigl[\overline{w}_i(X_i)\bigr]
  \;=\; \sum_{j=1}^\infty \lambda_j \sum_{i=1}^n \E\bigl[w_i^{(j)}(X_i)\bigr]
  \;\le\; \sum_{j=1}^\infty \lambda_j\,\frac{2^{-j}}{6}
  \;\le\; \frac{1}{24}.
  \label{eq:expected-pooling}
\end{equation}

\subsection{Sparse pooling across levels}

To relate each single-cutoff experiment \(\Pi_j\) to \(\Pi\), we first lower-bound the expectation of a product \(R = \prod_{i=1}^n (1 - W_i)\) for independent \(W_1,\ldots,W_n \in [0,1/2]\) on an arbitrary event \(E\). Even though a linear bound \(\E[\1_E\,R] \ge c\,\Pp(E)\) cannot hold uniformly in \(n\) (for instance, if each \(W_i\) equals \(1/2\) with probability \(1/n\) and \(0\) otherwise, then \(\E[\1_E\,R] = 2^{-n}\Pp(E)\) on the event \(E = \{W_1 = \cdots = W_n = 1/2\}\) of probability \(n^{-n}\)), bounding the negative moments of \(R\) gives an almost linear bound in \(\Pp(E)\).

\begin{lemma}[Conditional product bound]
\label{lem:conditional-product-bound}
Let \(W_1,\ldots,W_n \in [0,1/2]\) be independent random variables with \(\sum_{i=1}^n \E[W_i] \le 1/2\), and set \(R := \prod_{i=1}^n (1 - W_i)\). Then for every \(r > 0\) and every event \(E\),
\[
  \E[\1_E\,R] \;\ge\; \ee^{-2^r/r}\,\Pp(E)^{1 + 1/r}.
\]
\end{lemma}

\begin{proof}
Since \(z \mapsto (1 - z)^{-r}\) is convex on \([0,1/2]\), we have \((1 - z)^{-r} \le 1 + 2(2^r - 1)z \le 1 + 2^{r+1}z\) for all \(z \in [0,1/2]\). Applying this to \(W_i \in [0,1/2]\), taking expectations, and using \(1 + x \le \ee^x\) gives
\[
  \E\bigl[(1 - W_i)^{-r}\bigr]
  \;\le\; 1 + 2^{r+1}\,\E[W_i]
  \;\le\; \exp\bigl(2^{r+1}\,\E[W_i]\bigr).
\]
Multiplying across the independent variables \(W_1,\ldots,W_n\) and using \(\sum_{i=1}^n \E[W_i] \le 1/2\) gives
\[
  \E[R^{-r}] \;\le\; \exp\biggl(2^{r+1}\sum_{i=1}^n \E[W_i]\biggr) \;\le\; \ee^{2^r}.
\]

If \(\Pp(E) = 0\), the bound holds immediately. If \(\Pp(E) > 0\), Jensen's inequality for the convex function \(z \mapsto z^{-r}\) on \((0,\infty)\) gives
\[
  \E[R \mid E]^{-r}
  \;\le\; \E[R^{-r} \mid E]
  \;\le\; \frac{\E[R^{-r}]}{\Pp(E)}
  \;\le\; \frac{\ee^{2^r}}{\Pp(E)},
\]
so \(\E[R \mid E] \ge \ee^{-2^r/r}\,\Pp(E)^{1/r}\). Multiplying by \(\Pp(E)\) completes the proof.
\end{proof}

\begin{proof}[Proof of the {\thmref[universal guarantee]{thm:universal-guarantee}}]
Let \(\Pi\) be the \defref[universal experiment]{def:universal-experiment}, which depends only on \(P\). Given \(\mathbf{X} = (X_1,\ldots,X_n)\), the probability that every source reveals under \(\Pi\) is
\[
  R(\mathbf{X}) \;:=\; \prod_{i=1}^n \bigl(1 - \overline{w}_i(X_i)\bigr).
\]

\emph{Step 1: Relating \(\Pi\) to \(\Pi_j\).} Fix \(j \ge 1\), and let \(E := \bigl\{Y_{\Pi_j} \ge t_j,\ N \le 1\bigr\}\), where \(N\) is the number of pooling sources in \(\Pi_j\). Conditional on \(\mathbf{X}\), each of the \(n + 1\) disjoint cases of \(\{N \le 1\}\) under \(\Pi_j\) has a counterpart under \(\Pi\) that induces the same posterior mean, with probability at least \(\lambda_j R(\mathbf{X})\) times as large:
\begin{itemize}
\item \emph{Every source reveals (\(N = 0\)):} this occurs with probability at most \(1\) under \(\Pi_j\), and with probability \(R(\mathbf{X}) \ge \lambda_j R(\mathbf{X})\) under \(\Pi\).
\item \emph{Only source \(i\) pools (\(i \in [n]\)):} this occurs with probability at most \(w_i^{(j)}(X_i)\) under \(\Pi_j\), while under \(\Pi\) the counterpart where source \(i\) sends its level-\(j\) pool signal and every other source reveals occurs with probability \(\lambda_j w_i^{(j)}(X_i)\prod_{k \ne i}\bigl(1 - \overline{w}_k(X_k)\bigr) \ge \lambda_j R(\mathbf{X})\,w_i^{(j)}(X_i)\).
\end{itemize}
Since these \(n + 1\) counterparts under \(\Pi\) are disjoint, summing over those with posterior mean at least \(t_j\) gives \(\Pp(Y_\Pi \ge t_j \mid \mathbf{X}) \ge \lambda_j R(\mathbf{X})\,\Pp(E \mid \mathbf{X})\), and taking expectations over \(\mathbf{X} \sim P\) gives
\begin{equation}
  V_\Pi(t_j)
  \;\ge\; \lambda_j\,\E\bigl[R(\mathbf{X})\,\Pp(E \mid \mathbf{X})\bigr]
  \;=\; \lambda_j\,\E\bigl[\1_E\,R(\mathbf{X})\bigr].
  \label{eq:single-level-bound}
\end{equation}

\emph{Step 2: Guarantee for a fixed \(r \ge 1\).} Fix \(r \ge 1\). For every \(j \ge 1\), the \lemref[single-cutoff experiment]{lem:single-cutoff} with \(q = 2^{-j}\) gives \(\Pp(E) \ge \frac{1}{36}\,2^{-2j}\). Since the pooling probabilities \(W_i := \overline{w}_i(X_i) \in [0,1/2]\) are independent with \(\sum_{i=1}^n \E[W_i] \le 1/2\) by \cref{eq:expected-pooling}, applying the \lemref[conditional product bound]{lem:conditional-product-bound} to \cref{eq:single-level-bound} gives
\[
  V_\Pi(t_j)
  \;\ge\; \lambda_j\,\ee^{-2^r/r}\,\biggl(\frac{2^{-2j}}{36}\biggr)^{1 + 1/r}.
\]
Write \(\bigl(2^{-2j}\bigr)^{1 + 1/r} = 2^{j/r}\bigl(2^{-j}\bigr)^{2 + 3/r}\), and note that \(\ee^x \ge x^2/2\) for \(x = j\ln 2 / r\) gives
\[
  \lambda_j\,2^{j/r} \;=\; \frac{\exp(j\ln 2 / r)}{2j(j+1)} \;\ge\; \frac{(\ln 2)^2}{8r^2}.
\]
For any cutoff \(t \in \R\) with \(q := \Vjoint(t) > 0\), choosing \(j \ge 1\) with \(2^{-j} \ge q/2\) and \(t_j \ge t\) as in \cref{eq:dyadic-cutoff} therefore gives \(V_\Pi(t) \ge V_\Pi(t_j) \ge \frac{c_0}{r^2}\,\ee^{-2^r/r}\,\Vjoint(t)^{2 + 3/r}\) for an absolute constant \(c_0 > 0\). Applying the \lemref[cutoff reduction for a fixed experiment]{lem:fixed-exp-reduction} then gives
\begin{equation}
  V_\Pi(g) \;\ge\; \frac{c_0}{r^2}\,\ee^{-2^r/r}\,\Vjoint(g)^{2 + 3/r}
  \qquad\text{for every decision function } g \text{ and every } r \ge 1.
  \label{eq:universal-r-bound}
\end{equation}

\emph{Step 3: Optimizing \(r\).} Because \(\Pi\) does not depend on \(r\), we can choose \(r \ge 1\) in \cref{eq:universal-r-bound} separately for each \(g\). For any \(g\) with \(q := \Vjoint(g) > 0\), write \(L_q := \ln(\ee/q) \ge 1\) and choose \(r := \log_2(\ee + L_q) \ge 1\). Then \(2^r = \ee + L_q\) and \(q^{3/r} \ge \ee^{-3L_q/r}\), so both \(\ee^{-2^r/r}\) and \(q^{3/r}\) are at least \(\exp\Bigl(-O\bigl(L_q/\ln(\ee + L_q)\bigr)\Bigr)\), and absorbing \(1/r^2\) into the constant in the exponent gives the first bound. For any \(\varepsilon > 0\), taking \(r := \max\bigl\{1, 3/\varepsilon\bigr\}\) gives \(V_\Pi(g) \ge c_\varepsilon\,\Vjoint(g)^{2 + \varepsilon}\).
\end{proof}

\begin{remark}[Randomizing over product experiments]
The sub-polynomial factor \(\exp\Bigl(-\Theta\bigl(\ln(1/q)/\ln\ln(1/q)\bigr)\Bigr)\) in the \thmref[universal guarantee]{thm:universal-guarantee} is unavoidable for a single product experiment (\thmref[no universal quadratic bound]{thm:no-universal-quadratic}). Suppose instead that the designer can randomize over product experiments by drawing a single level \(j \ge 1\) with probability \(\frac{1}{j(j+1)}\) (equivalently, using shared randomness observed by the receiver) and having all sources run \(\Pi_j\). Then for every cutoff \(t\) with \(q = \Vjoint(t) > 0\), choosing \(j = \max\Bigl\{1, \bigl\lceil\log_2(1/q)\bigr\rceil\Bigr\}\) gives \(2^{-j} \ge q/2\) and \(t_j \ge t\), achieving probability of action at least
\[
  \frac{1}{j(j+1)}\,V_{\Pi_j}(t)
  \;\ge\; \frac{2^{-2j}}{36j(j+1)}
  \;\ge\; \frac{q^2}{144j(j+1)}
  \;\ge\; \frac{q^2}{576\ln^2(\ee/q)}.
\]
Since \(q \mapsto q^2/\ln^2(\ee/q)\) is nondecreasing and convex on \((0,1]\), the proof of the \lemref[cutoff reduction for a fixed experiment]{lem:fixed-exp-reduction} extends this \(\Omega\bigl(q^2/\ln^2(\ee/q)\bigr)\) guarantee to every decision function \(g\) with \(q = \Vjoint(g) > 0\), reducing the sub-polynomial loss to a logarithmic factor.
\end{remark}

\addtocontents{toc}{\protect\columnbreak}
\section{Tightness and limitations}
\label{sec:limitations}

We now turn to lower bounds and impossibility results. Except in \cref{sec:universal-bound}, which studies universal experiments, all of the negative results in this section hold even for a single known cutoff.

\subsection{Tightness of the quadratic bound}

We first show that the \thmref[quadratic guarantee for a known decision function]{thm:known-guarantee} is tight up to the constant factor, even for two sources and a known cutoff. The idea is to pair a coordinate \(X_1\) that is usually zero but occasionally very large with a coordinate \(X_2\) that lies just below the cutoff with probability \(q\): a joint experiment can pool the rare large realization of \(X_1\) selectively with the positive realization of \(X_2\) to raise the pool mean to the cutoff, whereas under a product experiment source \(1\) must pool without knowing whether \(X_2\) is positive.

\begin{theorem}[Two-source quadratic lower bound]
\label{thm:two-source-lower}
For every \(q \in (0,1)\) and \(\delta \in (0,1)\), there exist a two-source product prior \(P = P_1 \otimes P_2\) and a cutoff \(t \in \R\) such that \(\Vjoint(t) \ge q\) and \(\Vprod(t) \le (q + \delta)^2\).
\end{theorem}

\begin{proof}
Fix \(q \in (0,1)\) and \(\delta \in (0,1)\), and consider the cutoff \(t = 1\). Let \(X_1\) and \(X_2\) be independent with
\[
  X_1 \;=\; 1 + \frac{q}{\delta} \quad\text{with probability } \delta^2,
  \qquad\qquad
  X_2 \;=\; 1 - \delta \quad\text{with probability } q,
\]
and \(X_i = 0\) otherwise.

\emph{Joint value.} The event \(\{X_1 + X_2 > 0\}\) has probability \(q + (1 - q)\delta^2 \ge q\), and
\[
  \E[X_1 + X_2] \;=\; (q\delta + \delta^2) + q(1 - \delta) \;=\; q + \delta^2 \;\ge\; \Pp(X_1 + X_2 > 0).
\]
Since \(X_1 + X_2 = 0\) outside \(\{X_1 + X_2 > 0\}\), pooling \(\{X_1 + X_2 > 0\}\) induces posterior mean \(\E[X_1 + X_2]/\allowbreak\Pp(X_1 + X_2 > 0) \ge 1\), so \(\Vjoint(1) \ge q\).

\emph{Product experiments.} Under any product experiment, the posterior means \(Y_1\) and \(Y_2\) are independent with
\[
  Y_1 \;\ge\; 0, \qquad
  Y_2 \;\in\; [0, 1 - \delta], \qquad
  \E[Y_1] \;=\; q\delta + \delta^2, \qquad
  \E[Y_2] \;=\; q(1 - \delta).
\]
Whenever \(Y_1 + Y_2 \ge 1\), these constraints give
\[
  Y_1(Y_2 + \delta) \;\ge\; (1 - Y_2)(Y_2 + \delta) \;=\; \delta + Y_2(1 - \delta - Y_2) \;\ge\; \delta.
\]
Markov's inequality and the independence of \(Y_1\) and \(Y_2\) then give
\[
  \Pp(Y_1 + Y_2 \ge 1)
  \;\le\; \Pp\bigl(Y_1(Y_2 + \delta) \ge \delta\bigr)
  \;\le\; \frac{1}{\delta}\,\E[Y_1]\,\E[Y_2 + \delta]
  \;=\; (q + \delta)\bigl(q + (1 - q)\delta\bigr)
  \;\le\; (q + \delta)^2,
\]
so \(\Vprod(1) \le (q + \delta)^2\).
\end{proof}

\subsection{Tightness of the universal bound}
\label{sec:universal-bound}

We next show that no single product experiment can achieve a quadratic bound \(V_\Pi(t) \ge c\,\Vjoint(t)^2\) across all cutoffs with a constant \(c > 0\) independent of \(n\), and that the sub-polynomial factor in the \thmref[universal guarantee]{thm:universal-guarantee} is tight up to the constant in the exponent. The idea is to create a tension between two cutoffs: a low cutoff \(t_{\mathrm{lo}}\) at the prior mean, where the designer wants the sources to reveal nothing so that the sum stays at its mean, and a high cutoff \(t_{\mathrm{hi}}\) near \(n\), which requires \(n\) binary sources to send posterior means near \(1\) at the same time. Whenever a source sends a posterior mean near \(1\) with some probability, its posterior mean must fall below its prior mean the rest of the time, which pushes the sum below \(t_{\mathrm{lo}}\). Succeeding at \(t_{\mathrm{lo}}\) therefore limits how often the sources can send posterior means near \(1\), and multiplying this loss across all \(n\) independent sources reduces the probability of reaching \(t_{\mathrm{hi}}\) by an exponential factor in \(n\).

\begin{theorem}[No universal quadratic bound]
\label{thm:no-universal-quadratic}
For every integer \(n \ge 2\):
\begin{enumerate}[label=(\roman*)]
\item There are \(n + 1\) independent nonnegative coordinates \(X_0,X_1,\ldots,X_n\), each supported on two points, and two cutoffs \(t_{\mathrm{lo}} < t_{\mathrm{hi}}\) such that every product experiment \(\Pi\) satisfies
\[
  \min_{t \in \{t_{\mathrm{lo}}, t_{\mathrm{hi}}\}} \frac{V_\Pi(t)}{\Vjoint(t)^2} \;<\; \frac{\ln(n+1)}{n}.
\]
\item For every \(\gamma \in (0,1/2]\), there are \(n + 1\) independent nonnegative coordinates \(X_0,X_1,\ldots,X_n\), each supported on two points, and two cutoffs \(t_{\mathrm{lo}} < t_{\mathrm{hi}}\) with \(\Vjoint(t_{\mathrm{lo}}) = 1\) and \(Q := \Vjoint(t_{\mathrm{hi}}) \in (0,2^{-n})\) such that every product experiment \(\Pi\) with \(V_\Pi(t_{\mathrm{lo}}) \ge \gamma\) satisfies
\[
  V_\Pi(t_{\mathrm{hi}}) \;\le\; 5Q^2 \exp\biggl(-c_\gamma\,\frac{\ln(1/Q)}{\ln\ln(1/Q)}\biggr),
\]
where \(c_\gamma > 0\) depends only on \(\gamma\).
\end{enumerate}
\end{theorem}

\begin{proof}
\emph{Construction and joint values.}
Fix \(n \ge 2\). Given \(\mu \in (0,1/n)\), set \(q := \mu^n\). Let \(X_1,\ldots,X_n \sim \operatorname{Bernoulli}(\mu)\) be independent, and let \(X_0\) be independent of \((X_1,\ldots,X_n)\) with
\[
  X_0 \;=\; n + q^2 + q^{-1} \quad\text{with probability } q^4,
  \qquad\text{and}\qquad
  X_0 \;=\; 0 \quad\text{otherwise},
\]
so that \(\E[X_0] = q^3 + (n + q^2)q^4\). Define the cutoffs
\[
  t_{\mathrm{lo}} \;:=\; n\mu + \E[X_0] \;=\; \E\biggl[\sum_{i=0}^n X_i\biggr]
  \qquad\text{and}\qquad
  t_{\mathrm{hi}} \;:=\; n + q^2,
\]
and assume \(t_{\mathrm{lo}} < 1\) (so \(t_{\mathrm{lo}} < t_{\mathrm{hi}}\)).

The uninformative experiment gives \(\Vjoint(t_{\mathrm{lo}}) = 1\). At \(t_{\mathrm{hi}}\), the event \(A := \{X_0 > 0\} \cup \{X_1 = \cdots = X_n = 1\}\) has probability \(q^4 + (1 - q^4)q \ge q\) and
\[
  \E\Biggl[\1_A\biggl(\sum_{i=0}^n X_i - t_{\mathrm{hi}}\biggr)\Biggr]
  \;=\; q^4(q^{-1} + n\mu) - (1 - q^4)q \cdot q^2
  \;=\; n\mu q^4 + q^7
  \;>\; 0,
\]
so pooling \(A\) gives \(Q := \Vjoint(t_{\mathrm{hi}}) \ge q\).

\emph{Product experiments.}
Fix a product experiment \(\Pi\) with posterior means \(Y_0,Y_1,\ldots,Y_n\) and \(Y_\Pi = \sum_{i=0}^n Y_i\). Then \(Y_0 \ge 0\) with \(\E[Y_0] = \E[X_0]\), and \(Y_i \in [0,1]\) with \(\E[Y_i] = \mu\) for \(i \in [n]\). Since \(\E[Y_\Pi] = t_{\mathrm{lo}}\),
\begin{equation}
  \sum_{i=1}^n \E\bigl[(Y_i - t_{\mathrm{lo}})_+\bigr]
  \;\le\; \E\bigl[(Y_\Pi - t_{\mathrm{lo}})_+\bigr]
  \;=\; \E\bigl[(t_{\mathrm{lo}} - Y_\Pi)_+\bigr]
  \;\le\; t_{\mathrm{lo}}\bigl(1 - V_\Pi(t_{\mathrm{lo}})\bigr).
  \label{eq:low-cutoff-surplus}
\end{equation}
For each \(i \in [n]\), set \(Z_i := \frac{(Y_i - t_{\mathrm{lo}})_+}{1 - t_{\mathrm{lo}}} \in [0,1]\). For any \(\gamma \in (0,1)\) with \(V_\Pi(t_{\mathrm{lo}}) \ge \gamma\), the AM--GM inequality and \cref{eq:low-cutoff-surplus} give
\[
  \prod_{i=1}^n \E[Z_i]
  \;\le\; \biggl(\frac{1}{n}\sum_{i=1}^n \E[Z_i]\biggr)^n
  \;\le\; \biggl(\frac{t_{\mathrm{lo}}(1 - \gamma)}{n(1 - t_{\mathrm{lo}})}\biggr)^n.
\]

Now consider the high cutoff \(t_{\mathrm{hi}} = n + q^2\). On \(\{Y_\Pi \ge t_{\mathrm{hi}}\}\), rewriting \(Y_\Pi \ge n + q^2\) as \(Y_0 \ge \sum_{i=1}^n (1 - Y_i) + q^2\) and using \(Y_1,\ldots,Y_n \le 1\) and \(1 - Z_i \le \frac{1 - Y_i}{1 - t_{\mathrm{lo}}}\) gives both \(Y_0 \ge q^2\) and
\[
  1 - \prod_{i=1}^n Z_i
  \;\le\; \sum_{i=1}^n (1 - Z_i)
  \;\le\; \frac{1}{1 - t_{\mathrm{lo}}}\sum_{i=1}^n (1 - Y_i)
  \;\le\; \frac{Y_0}{1 - t_{\mathrm{lo}}}.
\]
Splitting \(1 = \bigl(1 - \prod_{i=1}^n Z_i\bigr) + \prod_{i=1}^n Z_i\) on \(\{Y_\Pi \ge t_{\mathrm{hi}}\}\) therefore gives
\[
  \1_{\{Y_\Pi \ge t_{\mathrm{hi}}\}}
  \;\le\; \frac{Y_0}{1 - t_{\mathrm{lo}}} + \frac{Y_0}{q^2}\prod_{i=1}^n Z_i.
\]
Taking expectations, using the independence of \(Y_0,Y_1,\ldots,Y_n\), and substituting \(Q \ge q\) and \(q^4 = q^3\mu^n\) with the bound on \(\prod_{i=1}^n \E[Z_i]\) above gives, whenever \(V_\Pi(t_{\mathrm{lo}}) \ge \gamma\),
\begin{equation}
  \frac{V_\Pi(t_{\mathrm{hi}})}{Q^2}
  \;\le\; \frac{\E[X_0]}{(1 - t_{\mathrm{lo}})q^2} + \frac{\E[X_0]}{q^4}\prod_{i=1}^n \E[Z_i]
  \;\le\; \frac{\E[X_0]}{(1 - t_{\mathrm{lo}})q^2} + \frac{\E[X_0]}{q^3}\biggl(\frac{t_{\mathrm{lo}}(1 - \gamma)}{n\mu(1 - t_{\mathrm{lo}})}\biggr)^n.
  \label{eq:two-cutoff-bound}
\end{equation}

\emph{Completing the proof.}
For Part~(i), set \(\gamma := \frac{\ln(n+1)}{n}\). As \(\mu \downarrow 0\), we have \(\E[X_0] \sim q^3\) and \(t_{\mathrm{lo}} \sim n\mu\), so the right-hand side of \cref{eq:two-cutoff-bound} converges to \((1 - \gamma)^n < \ee^{-n\gamma} = \frac{1}{n+1} < \gamma\). For small enough \(\mu > 0\), every \(\Pi\) with \(V_\Pi(t_{\mathrm{lo}})/\Vjoint(t_{\mathrm{lo}})^2 = V_\Pi(t_{\mathrm{lo}}) \ge \gamma\) therefore has \(V_\Pi(t_{\mathrm{hi}})/Q^2 < \gamma\).

For Part~(ii), given \(\gamma \in (0,1/2]\), set \(\mu := \frac{\gamma}{4n}\). Since \(\sum_{i=1}^n X_i \le n\), we have \(\sum_{i=0}^n X_i > n\) only when \(X_0 > 0\), so \(\E\Bigl[\bigl(\sum_{i=0}^n X_i - n\bigr)_+\Bigr] = q^4\bigl(q^{-1} + q^2 + n\mu\bigr) < 2q^3\). Because the top-\(Q\) pool of \(\sum_{i=0}^n X_i\) has mean at least \(t_{\mathrm{hi}} = n + q^2\), this gives
\[
  Q \;\le\; \frac{1}{q^2}\,\E\Biggl[\biggl(\sum_{i=0}^n X_i - n\biggr)_+\Biggr] \;<\; 2q \;<\; 2^{-n}.
\]
Moreover, \(q = \mu^n\) and \(n\mu = \gamma/4\) give \(\E[X_0] < 2q^3 \le \frac{\gamma}{4}n\mu\), so \(t_{\mathrm{lo}} = n\mu + \E[X_0] \le (1 + \gamma/4)n\mu < \gamma/2\) and hence
\[
  \frac{t_{\mathrm{lo}}(1 - \gamma)}{n\mu(1 - t_{\mathrm{lo}})}
  \;\le\; \frac{(1 + \gamma/4)(1 - \gamma)}{1 - \gamma/2}
  \;\le\; \ee^{-\gamma/4}.
\]
Substituting these bounds into \cref{eq:two-cutoff-bound} gives
\[
  \frac{V_\Pi(t_{\mathrm{hi}})}{Q^2}
  \;\le\; 3q + 2\ee^{-\gamma n / 4}
  \;\le\; 5\ee^{-\gamma n / 4}
  \;\le\; 5\exp\biggl(-c_\gamma\,\frac{\ln(1/Q)}{\ln\ln(1/Q)}\biggr)
\]
for a constant \(c_\gamma > 0\) depending only on \(\gamma\), since \(Q \in [q, 2q)\) and \(q = \bigl(\gamma/(4n)\bigr)^n\) give \(n = \Theta_\gamma\bigl(\ln(1/Q)/\ln\ln(1/Q)\bigr)\).
\end{proof}

\subsection{Discontinuity at one}

When \(\Vjoint(t) = 1\), we have \(t \le \E[X]\), so the uninformative experiment achieves \(\Vprod(t) = 1\). Moreover, for any fixed \(n\), \propref[greedy coordinatewise pooling]{prop:greedy} gives \(\Vprod(t) \ge \Vjoint(t)^n \to 1\) as \(\Vjoint(t) \uparrow 1\). However, this convergence is not uniform in \(n\). Intuitively, a joint experiment can pool many rare positive values so that the receiver acts with probability \(1 - \varepsilon\), whereas the sum of \(n\) independent nonnegative posterior means falls below its mean plus the largest coordinate mean with probability at least \(1/\ee\). This tight small-deviation inequality was conjectured by \citelink{Uriel Feige}{Feige}~\cite{Feige}, who proved it with \(1/13\) in place of \(1/\ee\). The sharp constant \(1/\ee\) was proved independently by \citelink{Weibo Fu, Yanjun Han, Guanyang Wang, Jun Yan, Peng Zhang, and Zhengqing Zhou}{FuEtAl}~\cite{FuEtAl} and by \citelink{Zipei Nie and Jiaye Wei}{NieWei}~\cite{NieWei}.

\begin{theorem}[Feige's small-deviation inequality]
\label{thm:feige}
Let \(W_1,\ldots,W_n\) be independent nonnegative random variables with \(\E[W_i] \le 1\) for every \(i \in [n]\). Then
\[
  \Pp\biggl(\sum_{i=1}^n W_i \;<\; \sum_{i=1}^n \E[W_i] + 1\biggr) \;\ge\; \frac{1}{\ee}.
\]
\end{theorem}

\begin{proposition}[Discontinuity at one]
\label{prop:discontinuity-at-one}
For every \(\varepsilon \in (0,1)\), there exist a product prior and a cutoff \(t\) with \(\Vjoint(t) \ge 1 - \varepsilon\) and \(\Vprod(t) \le 1 - 1/\ee\).
\end{proposition}

\begin{proof}
Choose an integer \(n \ge (1 - \varepsilon)/\varepsilon\), and let \(X_1,\ldots,X_n\) be independent with
\[
  X_i \;=\; n + 1 \quad\text{with probability } \frac{1}{n+1},
  \qquad\text{and}\qquad
  X_i \;=\; 0 \quad\text{otherwise},
\]
so that \(\E[X_i] = 1\) and \(\E[X] = n\).

\emph{Joint value.} Since every positive realization of \(X\) is at least \(n + 1\), Markov's inequality gives
\[
  \Pp(X > 0) \;\le\; \frac{\E[X]}{n+1} \;=\; \frac{n}{n+1}.
\]
Pooling \(\{X > 0\}\) together with enough of \(\{X = 0\}\) to get total probability \(n/(n+1)\) includes all positive realizations of \(X\), so its conditional mean is \(\E[X]/\bigl(n/(n+1)\bigr) = n + 1\). Thus \(\Vjoint(n + 1) \ge n/(n+1) \ge 1 - \varepsilon\) by the choice of \(n\).

\emph{Product experiments.} Under any product experiment, the posterior means \(Y_1,\ldots,Y_n\) are independent and nonnegative with mean \(1\), so \thmref{thm:feige} gives \(\Pp\bigl(\sum_{i=1}^n Y_i < n + 1\bigr) \ge 1/\ee\), and hence \(\Vprod(n + 1) \le 1 - 1/\ee\).
\end{proof}

\subsection{The role of independence}

Suppose that the sources still signal independently, each observing only its own coordinate, but that the coordinates are correlated under the prior. The receiver then updates on all signals jointly, and her posterior mean of \(X\) need not equal \(\sum_{i=1}^n \E[X_i \mid S_i]\). The next proposition shows that \(\Vprod(t)\) can be arbitrarily small even when \(\Vjoint(t) = 1/2\), already with two sources and three states.

\begin{proposition}[Impossibility under correlated priors]
\label{prop:correlated-prior-impossibility}
For every integer \(K \ge 2\), there is a prior on \((X_1, X_2)\) with correlated coordinates such that \(\Vjoint(0) = 1/2\) while \(\Vprod(0) < 3/(2K)\).
\end{proposition}

\begin{proof}
Consider the prior on \((X_1, X_2)\) that assigns probabilities
\[
  \frac{1}{2(K+1)} \;\text{ to }\; (0, K),
  \qquad
  \frac{K}{2(K+1)} \;\text{ to }\; (K - 1, -K),
  \qquad
  \frac{1}{2} \;\text{ to }\; (0, -K),
\]
so that the payoffs \(X_1 + X_2\) in these states are \(K\), \(-1\), and \(-K\), respectively.

\emph{Joint value.} Pooling the first two states gives probability \(\frac{1}{2(K+1)} + \frac{K}{2(K+1)} = \frac{1}{2}\) and expected payoff \(\frac{K}{2(K+1)} - \frac{K}{2(K+1)} = 0\), so \(\Vjoint(0) \ge \frac{1}{2}\). Conversely, in any pool on which the receiver acts, each unit of mass on the positive state \((0, K)\) can make up for at most \(K\) units of mass on the other two states, whose payoffs are at most \(-1\). Hence every such pool has probability at most \((K + 1) \cdot \frac{1}{2(K+1)} = \frac{1}{2}\).

\emph{Product experiments.} A joint experiment pools \((0, K)\) with \((K - 1, -K)\) while excluding \((0, -K)\). Under a product experiment \(\Pi\), however, source \(1\) cannot distinguish \((0, K)\) from \((0, -K)\) (both have \(X_1 = 0\)), and source \(2\) cannot distinguish \((K - 1, -K)\) from \((0, -K)\) (both have \(X_2 = -K\)).

Specifically, since \((K - 1, -K)\) has negative payoff \(-1\), the receiver can act given \((S_1, S_2)\) only if her conditional expected payoff across \((0, K)\) and \((0, -K)\) is nonnegative. Because source \(1\) observes \(X_1 = 0\) in both of these states, \(S_1 \sim \pi_1(\cdot \mid 0)\) is independent of \((X_2, S_2)\) conditional on \(X_1 = 0\), so the receiver acts only on the event
\[
  E \;:=\; \bigl\{\E[X_1 + X_2 \mid S_2,\, X_1 = 0] \ge 0\bigr\},
\]
which depends only on \(S_2 \sim \pi_2(\cdot \mid X_2)\) and thus has the same conditional probability \(\Pp(E \mid X_2 = -K)\) in both \((K - 1, -K)\) and \((0, -K)\). Since \(\{X_1 = 0\}\) consists of \((0, K)\) and \((0, -K)\), which have payoffs \(K\) and \(-K\) and prior probabilities \(\frac{1}{2(K+1)}\) and \(\frac{1}{2}\), the definition of \(E\) gives
\[
  0 \;\le\; \E\bigl[\1_{E \cap \{X_1 = 0\}}(X_1 + X_2)\bigr]
  \;=\; \frac{K}{2(K+1)}\,\Pp(E \mid X_2 = K) - \frac{K}{2}\,\Pp(E \mid X_2 = -K),
\]
so \(\Pp(E \mid X_2 = -K) \le \frac{1}{K+1}\). Since the receiver acts only on \(E\), we get
\[
  V_\Pi(0)
  \;\le\; \Pp(E)
  \;\le\; \Pp(X_2 = K) + \Pp(E \mid X_2 = -K)
  \;\le\; \frac{1}{2(K+1)} + \frac{1}{K+1}
  \;=\; \frac{3}{2(K+1)}
  \;<\; \frac{3}{2K}. \qedhere
\]
\end{proof}

\subsection{The role of additivity}

Finally, we keep the prior independent and ask whether our guarantees extend to a broader class of monotone payoffs. A function \(f\colon \{0,1\}^n \to [0,\infty)\) is \emph{XOS} (fractionally subadditive) if it is a pointwise maximum of nonnegative additive functions, that is, \(f(x_1,\ldots,x_n) = \max_\ell \sum_{i=1}^n w_{\ell,i} x_i\) with \(w_{\ell,i} \ge 0\). Let \(X_1,\ldots,X_n \sim \operatorname{Bernoulli}(1/2)\) be independent, and write \(\mathbf{X} := (X_1,\ldots,X_n)\). When the receiver's payoff from acting is \(f(\mathbf{X})\), we write \(\Vjoint^f(t)\) and \(\Vprod^f(t)\) for the supremum probabilities that \(\E\bigl[f(\mathbf{X}) \mid \text{signals}\bigr] \ge t\) under joint and product experiments, respectively.

Let \(n = 2k \ge 2\) be even, write \(X := \sum_{i=1}^n X_i\), and define
\[
  f(\mathbf{X}) \;:=\; \min\{X, k\} + 8^{-n}\,\1_{\{X = n\}},
  \qquad
  t \;:=\; k + 2 \cdot 16^{-n}.
\]
The payoff is capped at \(k = n/2\), with a small bonus of \(8^{-n}\) when \(X_1 = \cdots = X_n = 1\). A joint experiment can pool the upper half of the prior, on which \(X \ge k\), so the capped payoff equals \(k\) throughout the pool and the rare bonus raises the pool mean to \(t\). A product experiment, by contrast, cannot pool the rare bonus at \(X = n\) with a large set of states without also including \(\{X < k\}\), and therefore induces action with only exponentially small probability.

\begin{proposition}[Impossibility for XOS payoffs]
\label{prop:xos-impossibility}
For every even integer \(n = 2k \ge 2\), the function \(f\) is nonnegative, monotone, symmetric, and XOS, and
\[
  \Vjoint^f(t) \;=\; \frac{1}{2},
  \qquad\text{whereas}\qquad
  \Vprod^f(t) \;\le\; \ee^{-n/128}.
\]
\end{proposition}

\begin{proof}
\emph{XOS representation.} Nonnegativity, monotonicity, and symmetry are clear; for the XOS property, note that
\[
  f(\mathbf{X}) \;=\; \max\biggl\{\max_{I \subseteq [n],\ |I| = k}\sum_{i \in I} X_i,\ \frac{k + 8^{-n}}{n}\,X\biggr\}.
\]

\emph{Joint value.} Pooling all states with \(X > k\) together with half of \(\{X = k\}\) gives a pool of probability \(1/2\) whose conditional mean payoff is \(k + 8^{-n}2^{-n}/(1/2) = t\). Conversely, since \(\E\Bigl[\bigl(f(\mathbf{X}) - k\bigr)_+\Bigr] = 16^{-n}\), any joint experiment induces action with probability at most \(16^{-n}/(t - k) = 1/2\).

\emph{Product experiments.} Fix a product experiment, and write \(Y_i := \Pp(X_i = 1 \mid S_i)\). These posterior means are independent, lie in \([0,1]\), and have mean \(1/2\). Conditional on the signals, the coordinates remain independent with \(X_i \sim \operatorname{Bernoulli}(Y_i)\). Since \(f(\mathbf{X}) = k - (k - X)_+ + 8^{-n}\,\1_{\{X = n\}}\), the action condition \(\E\bigl[f(\mathbf{X}) \mid \text{signals}\bigr] \ge t = k + 2 \cdot 16^{-n}\) becomes
\[
  8^{-n}\prod_{i=1}^n Y_i - \E\bigl[(k - X)_+ \mid \text{signals}\bigr] \;\ge\; 2 \cdot 16^{-n}.
\]

We claim that at least \(k = n/2\) posterior means \(Y_i\) exceed \(7/8\) whenever the receiver acts. Indeed, if \(Y_i \le 7/8\) (so \(1 - Y_i \ge Y_i/7\)) for a set \(I \subseteq [n]\) of \(k + 1\) sources, the state with \(X_i = 0\) for \(i \in I\) and \(X_i = 1\) for \(i \notin I\) has \(k - X = 1\) and posterior probability \(\prod_{i \in I}(1 - Y_i)\prod_{i \notin I} Y_i \ge 7^{-(k+1)}\prod_{i=1}^n Y_i \ge 8^{-n}\prod_{i=1}^n Y_i\), contradicting the inequality above.

Whenever the receiver acts, choose \(k\) sources with \(Y_i > 7/8\). Since their posterior means have product at most \(1\) and \(8^{-n}\prod_{i=1}^n Y_i \ge 2 \cdot 16^{-n}\), the product of the other \(k\) posterior means is at least \(\prod_{i=1}^n Y_i \ge 2^{1-n}\), so by the AM--GM inequality the other \(k\) posterior means have sum at least \(k\bigl(2^{1-n}\bigr)^{1/k} = \frac{k}{4}\,2^{1/k} > \frac{k}{4}\), giving
\[
  \sum_{i=1}^n Y_i \;>\; \frac{7k}{8} + \frac{k}{4} \;=\; \frac{9n}{16} \;=\; \frac{n}{2} + \frac{n}{16}.
\]
Since \(Y_1,\ldots,Y_n \in [0,1]\) are independent with mean \(1/2\), Hoeffding's inequality gives
\[
  \Vprod^f(t) \;\le\; \Pp\biggl(\sum_{i=1}^n Y_i \;>\; \frac{n}{2} + \frac{n}{16}\biggr) \;\le\; \ee^{-n/128}. \qedhere
\]
\end{proof}

\section{Concluding remarks}

We have shown that persuasion by parts always achieves at least a constant times the square of the joint value for any number of sources and any product prior, and that a single universal experiment achieves every exponent \(2 + \varepsilon\) simultaneously across all decision functions. We conclude with a few open questions and directions for future work.

\paragraph{Closing the gaps.}
For a known cutoff, let \(\rho(q)\) be the infimum of \(\Vprod(t)\) over all instances \((P, t)\) with \(\Vjoint(t) \ge q\). Our bounds (\thmref[quadratic guarantee for a known decision function]{thm:known-guarantee}, \thmref[two-source quadratic lower bound]{thm:two-source-lower}, and \propref[discontinuity at one]{prop:discontinuity-at-one}) place \(\rho(q)\) in \(\Bigl[\frac{1}{36}\,q^2,\, \min\bigl\{q^2, 1 - 1/\ee\bigr\}\Bigr]\) for every \(q \in (0,1)\). A natural open problem is to close this constant-factor gap: in particular, is \(\rho(q) = \bigl(1 - o(1)\bigr)\,q^2\) as \(q \to 0\), and what is \(\lim_{q \uparrow 1} \rho(q)\)?

\paragraph{Beyond independence, additivity, and a single sum.}
The \propref[impossibility under correlated priors]{prop:correlated-prior-impossibility} and the \propref[impossibility for XOS payoffs]{prop:xos-impossibility} show that neither independence of the prior nor additivity of the payoff can be dropped in general. Still, one may ask whether guarantees independent of \(n\) hold under natural restrictions on the correlation across coordinates, or for monotone submodular payoffs, which lie strictly between additive and XOS functions. Even for additive payoffs, it would be interesting to understand what product experiments can achieve when persuading multiple receivers who place different weights on the coordinates.

\paragraph{Communication, prior knowledge, and incentives.}
Finally, our model assumes that a single designer knows the product prior \(P = \bigotimes_{i=1}^n P_i\) and chooses simultaneous signaling rules for all sources. Relaxing any of these assumptions leads to interesting questions. For example, if sources report sequentially rather than simultaneously, later sources can condition their signals on earlier reports and adjust their pools when earlier posterior means are low; how much of the joint value does sequential signaling recover? Similarly, if each distribution \(P_i\) is known only to source \(i\) (or only through samples), how much coordination across sources is needed to design near-optimal signaling rules? And what happens when the sources have their own incentives (for instance, separate certifiers selling component tests in a market) rather than following signaling rules chosen by a single designer?

\unnumberedsection{Acknowledgments}

This work was carried out while Kangning Wang was visiting Shanghai University of Finance and Economics. We thank Guanyang Wang, Jinzhao Wu, and Peng Zhang for helpful discussions.

\unnumberedsection{AI methodology and disclosure}

This project began in May 2026. At the time, we (the human authors) explored settings around ``simple'' persuasion rules when a receiver's payoff is the sum of independent coordinates. We soon arrived at our setting and the conjecture that \(\Vjoint(t) = 1 - \varepsilon\) might imply \(\Vprod(t) \ge \varepsilon\) for small enough \(\varepsilon > 0\). Neither our initial human attempts nor an extended early exploration with GPT-5.6 Sol Pro resolved this conjecture. (It now follows from the \thmref[quadratic guarantee for a known decision function]{thm:known-guarantee}.)

Over the course of the project, we used OpenAI's GPT-5.6 Sol and GPT-6 Astra extensively, and Gemini 4 Argon and Claude Opus 5.5 to a lesser extent, to develop the proofs in this paper. In the early stages, we supplied some of our own proof ideas to these AI systems. However, we think it is unlikely that those ideas played a meaningful role in the AI systems finding the proofs that eventually worked. As the project progressed, the quantitative bounds improved from exponential and quasi-polynomial to polynomial; these improvements mostly relied on KL-divergence probability bounds connecting the optimal joint pool with a product experiment. We went through a version of these earlier proofs and extracted some interesting technical ideas, but the core arguments were long and tedious without much intuitive structure as far as we could tell. Nevertheless, we were quite impressed by the technical prowess of these AI tools. The KL-divergence arguments gave a polynomial tradeoff but seemed unable to reach the optimal exponent. We were pleasantly surprised when later attempts gave a quadratic bound with a much simpler proof.

Beyond our early attempts, the human authors' technical role consisted of formulating and directing the research questions, iteratively selecting, restructuring, and cleaning up the proof ideas, and sometimes replacing proof steps with our own arguments. In our (anecdotal) experience, cleaning up AI-generated proofs and feeding the simplified versions back into later prompts did seem to help (though, of course, we have no control group for this).

We and the AI systems searched the literature together. For the writing, we initially worked with GPT-6 Astra, and later with Gemini 4 Argon and Claude Opus 5.5 as they became available and we became aware of their strengths. The writing process largely consisted of the human authors commenting on AI-drafted text and providing detailed feedback on the structure and presentation choices over several passes. It felt much like closely advising a student who is technically very strong, but whose writing habits are quite different from our own.

The authors verified and take full responsibility for all content in the paper.

\paperbibliography{references}{references}

\appendix
\crefalias{section}{appendix}

\section{Polynomial-time construction}
\label{sec:computation}

Suppose that each distribution \(P_i\) is given explicitly as a finite list of rational values and probabilities with total binary encoding length \(L\). Our constructions in \cref{sec:known,sec:universal} use the pool means \(m_X(s)\) and the quantiles of \(X_{-i}\). Although the exact distribution of \(X\) can have exponentially large support and computing its upper-tail probabilities is \(\#\mathrm{P}\)-hard, we can approximate these probabilities within any factor \(1 + \delta\) in time polynomial in \(L\) and \(1/\delta\) by standard dynamic programming. Plugging these approximate tails into the constructions of \cref{sec:known,sec:universal} over a polynomial-size geometric grid of cutoff levels implements both in deterministic polynomial time; the proof below simply repeats the main-body analysis while tracking the \(1 + \delta\) factors.

\begin{theorem}[Polynomial-time construction]
\label{thm:polytime}
For any product prior \(P = \bigotimes_{i=1}^n P_i\) given explicitly as above with encoding length \(L\):
\begin{enumerate}[label=(\roman*)]
\item Given \(\delta \in (0,1)\) and a cutoff \(t \in \R\), or a decision function \(g\) whose values at rational points are rational and computable in polynomial time, one can construct in deterministic time polynomial in \(L\) and \(1/\delta\) a product experiment \(\Pi\) with
\[
  V_\Pi(t) \;\ge\; \frac{1 - \delta}{36}\,\Vjoint(t)^2
  \qquad\text{or}\qquad
  V_\Pi(g) \;\ge\; \frac{1 - \delta}{144}\,\Vjoint(g)^2,
  \quad\text{respectively.}
\]
\item One can construct in deterministic time polynomial in \(L\) a single product experiment \(\Pi\) that satisfies the \thmref[universal guarantee]{thm:universal-guarantee}.
\end{enumerate}
\end{theorem}

\begin{lemma}[Simultaneous tail approximation]
\label{lem:tail-approximation}
Let \(W\) be the sum of any subset of the coordinates, and let \(\delta \in (0,1]\). One can compute in deterministic time polynomial in \(L\) and \(1/\delta\) a finitely supported rational random variable \(Z\) such that
\begin{equation}
  \frac{1}{1 + \delta}\,\Pp(W \ge x) \;\le\; \Pp(Z \ge x) \;\le\; \Pp(W \ge x)
  \qquad\text{for all } x \in \R,
  \label{eq:tail-sandwich}
\end{equation}
and hence \(m_W\bigl((1 + \delta)q\bigr) \le m_Z(q) \le m_W(q)\) for all \(q \in \bigl(0, \frac{1}{1 + \delta}\bigr]\).
\end{lemma}

\begin{proof}
Add the coordinates of \(W\) one at a time by convolving their distributions; to keep the support small, after each step round every positive upper-tail probability \(\Pp(\cdot \ge x)\) down to the nearest power \(\beta^{-k}\) of \(\beta := 1 + \frac{\delta}{2n}\) (keeping only the largest support point for each \(\beta^{-k}\)). Since positive probabilities are at least \(2^{-L}\), each intermediate distribution is supported on \(O(nL/\delta)\) rational points of polynomial bit length.

Convolving with an independent coordinate averages upper-tail probabilities and therefore preserves multiplicative bounds on \(\Pp(\cdot \ge x)\), while each rounding step loses at most a factor of \(\beta\). After at most \(n\) steps, the resulting random variable \(Z\) loses at most a factor of \(\beta^n \le \ee^{\delta/2} \le 1 + \delta\), giving \cref{eq:tail-sandwich}. Hence,
\[
  F_W^{-1}\bigl(1 - (1 + \delta)u\bigr) \;\le\; F_Z^{-1}(1 - u) \;\le\; F_W^{-1}(1 - u)
  \qquad\text{for all } u \in \Bigl(0,\frac{1}{1 + \delta}\Bigr),
\]
and averaging over \(u \in (0,q)\) gives \(m_W\bigl((1 + \delta)q\bigr) \le m_Z(q) \le m_W(q)\).
\end{proof}

\begin{proof}[Proof of the {\thmref[polynomial-time construction]{thm:polytime}}]
Given \(\eta \in (0,1]\), apply the \lemref[simultaneous tail approximation]{lem:tail-approximation} with parameter \(\eta\) to compute tail approximations \(Z\) for \(X\) and \(Z_{-i}\) for each \(X_{-i}\). For any \(q \in \bigl(0,\frac{1}{1 + \eta}\bigr]\), define the product experiment \(\Pi(q)\) as in the \lemref[single-cutoff experiment]{lem:single-cutoff} at cutoff \(t_q := m_Z(q)\), except with targets and pooling probabilities
\[
  h_i \;:=\; t_q - F_{Z_{-i}}^{-1}\Bigl(1 - \frac{q}{3(1 + \eta)}\Bigr)
  \qquad\text{and}\qquad
  s_i \;:=\; \frac{q}{6(1 + \eta)}\frac{b_i}{B}
\]
(pooling when \(B \ge \frac{q}{6(1 + \eta)}\), with \(b_i\) and \(B = \sum_{i=1}^n b_i\) computed directly from \(P_1,\ldots,P_n\)).

By \cref{eq:tail-sandwich}, we have \(m_X\bigl((1 + \eta)q\bigr) \le t_q \le m_X(q)\), \(\Pp(X_{-i} > t_q - h_i) \le q/3\), and \(\Pp(X_{-i} \ge t_q - h_i) \ge \frac{q}{3(1 + \eta)}\). The proofs of the \lemref[capacity lemma]{lem:capacity} and the \lemref[single-cutoff experiment]{lem:single-cutoff} therefore give
\begin{equation}
  \E[N] \;\le\; \frac{q}{6(1 + \eta)}
  \qquad\text{and}\qquad
  \Pp\bigl(Y_{\Pi(q)} \ge t_q,\ N \le 1\bigr) \;\ge\; \frac{q^2}{36(1 + \eta)^2}.
  \label{eq:polytime-level-guarantee}
\end{equation}
Moreover, any cutoff \(t \in \R\) with \(\Vjoint(t) > 0\) has \(\Vjoint(t) \ge 2^{-L}\), and \(t_q \ge m_X\bigl((1 + \eta)q\bigr) \ge t\) whenever \((1 + \eta)q \le \Vjoint(t)\).

For Part~(i), set \(\eta := \delta/6\), consider the grid \(q_k := (1 + \eta)^{-k}\) for \(1 \le k \le K := \bigl\lceil L/\log_2(1 + \eta)\bigr\rceil + 1\), and output \(\Pi := \Pi(q_{k^*})\) for an index \(k^*\) maximizing \(g(t_{q_k})\,q_k^2\) (with \(g = \1_{\{\cdot \ge t\}}\) for a known cutoff \(t\)). Every \(t \in \R\) with \(\Vjoint(t) > 0\) is covered by some \(q_k \in \bigl[\frac{\Vjoint(t)}{(1 + \eta)^2},\, \frac{\Vjoint(t)}{1 + \eta}\bigr]\), for which \(t_{q_k} \ge t\); by \cref{eq:polytime-level-guarantee} and the choice of \(k^*\),
\[
  V_\Pi(g)
  \;\ge\; \frac{g(t_{q_{k^*}})\,q_{k^*}^2}{36(1 + \eta)^2}
  \;\ge\; \frac{g(t_{q_k})\,q_k^2}{36(1 + \eta)^2}
  \;\ge\; \frac{g(t)\,\Vjoint(t)^2}{36(1 + \eta)^6}
  \;\ge\; \frac{1 - \delta}{36}\,g(t)\,\Vjoint(t)^2.
\]
Since this holds for all \(t \in \R\), the proof of the \lemref[cutoff reduction for a known decision function]{lem:known-g-reduction} gives \(V_\Pi(g) \ge \frac{1 - \delta}{144}\,\Vjoint(g)^2\).

For Part~(ii), set \(\eta := 1\) and mix \(\Pi(2^{-j})\) for \(1 \le j \le L + 1\) with weights \(\lambda_j = \frac{1}{2j(j+1)}\) as in the \defref[universal experiment]{def:universal-experiment}; Part~(ii) then follows from the proof of the \thmref[universal guarantee]{thm:universal-guarantee}.
\end{proof}

\end{document}

%% file: macros.tex
\usepackage[T1]{fontenc}
\usepackage[margin=1in]{geometry}
\usepackage{amsmath,amssymb,amsthm}
\usepackage{etoolbox}
\usepackage{libertinus}
\usepackage{libertinust1math}
\DeclareSymbolFont{cmlargesymbols}{OMX}{cmex}{m}{n}
\DeclareMathSymbol{\intop}{\mathop}{cmlargesymbols}{"52}
\usepackage{enumitem}
\usepackage{microtype}
\usepackage{titlesec}
\usepackage{multicol}
\usepackage[skins]{tcolorbox}
\usetikzlibrary{arrows.meta}

\definecolor{CitationBlue}{HTML}{205E9D}
\definecolor{InternalLink}{HTML}{7A3E65}
\definecolor{URLTeal}{HTML}{1B645B}
\definecolor{ContentsGray}{HTML}{546680}
\definecolor{ContentsHeading}{HTML}{40546F}

\definecolor{StatementAccent}{HTML}{2C5286}
\colorlet{DefinitionAccent}{URLTeal}
\definecolor{ExplanatoryAccent}{HTML}{785135}
\colorlet{ActiveStatementAccent}{StatementAccent}
\newcommand{\statementheadingfont}{\rmfamily\bfseries\upshape}

\colorlet{RevealAccent}{ExplanatoryAccent}
\colorlet{PoolAccent}{DefinitionAccent}

\usepackage{hyperref}
\hypersetup{
  colorlinks=true,
  linkcolor=InternalLink,
  citecolor=CitationBlue,
  urlcolor=URLTeal,
  linktoc=all,
  pdftitle={Persuasion by Parts},
  pdfauthor={Shuyuan Jia, Kangning Wang, and Renzhe Xu},
  pdfsubject={Bayesian persuasion under product information structures},
  pdfkeywords={Bayesian persuasion, independent information, economics and computation, probability},
  pdflang={en-US}
}
\newcommand{\citelink}[2]{%
  \hyperlink{cite.#2}{\textcolor{CitationBlue}{#1}}%
}
\usepackage[all]{hypcap}
\usepackage[capitalise,noabbrev,nameinlink]{cleveref}

\crefformat{equation}{#2Eq.~(#1)#3}
\crefmultiformat{equation}{#2Eqs.~(#1)#3}{ and~#2(#1)#3}{, #2(#1)#3}{, and~#2(#1)#3}

\newtheoremstyle{cardupright}
  {0pt}{0pt}{\upshape}{0pt}{\normalfont}{}{0pt}{\relax}

\numberwithin{equation}{section}
\theoremstyle{cardupright}
\newtheorem{theorem}{Theorem}[section]
\newtheorem{proposition}[theorem]{Proposition}
\newtheorem{lemma}[theorem]{Lemma}
\newtheorem*{introtheorem}{Main theorem}
\newtheorem{definition}[theorem]{Definition}
\newtheorem{example}[theorem]{Example}
\newtheorem{remark}[theorem]{Remark}

\newif\ifStatementNumbered
\tcbset{
  statementbox/.style={
    enhanced,
    colback=white,colframe=ActiveStatementAccent!35,coltitle=ActiveStatementAccent,
    colbacktitle=white,titlerule=0pt,
    boxrule=0.4pt,rounded corners,arc=3mm,boxsep=0pt,
    left=11pt,right=11pt,lefttitle=11pt,righttitle=96pt,
    top=5pt,bottom=9pt,toptitle=5pt,bottomtitle=2pt,
    fonttitle=\statementheadingfont\fontsize{9.5}{11}\selectfont,
    before title={\strut},
    before upper={%
      \setlength{\parindent}{0pt}%
      \setlength{\parskip}{0.55em plus 0.1em minus 0.1em}%
    },
    before skip=16pt,after skip=12pt,
    overlay unbroken={%
      \node[rectangle,rounded corners=10pt,fill=ActiveStatementAccent!14,
        draw=ActiveStatementAccent!55,line width=0.4pt,
        minimum width=82pt,minimum height=20pt,inner sep=0pt,outer sep=0pt]
        at ([xshift=-46pt,yshift=-3pt]frame.north east) {%
          \textcolor{ActiveStatementAccent}{%
            {\statementheadingfont\fontsize{7}{8}\selectfont\StatementBadgeKind}%
            \ifStatementNumbered\hspace{4pt}%
              {\statementheadingfont\fontsize{9.5}{11}\selectfont\StatementBadgeNumber}%
            \fi}%
        };
    },
  }
}
\makeatletter
\NewSocketPlug{hyp/raisedlink}{statementbox}{%
  \dimen@=\prevdepth
  \vbox to 0pt{\kern-36pt\hbox{#1}\vss}%
  \prevdepth=\dimen@
}
\makeatother
\newcommand{\wrapstatement}[4]{%
  \csletcs{statementoriginal#1}{#1}%
  \csletcs{endstatementoriginal#1}{end#1}%
  \renewenvironment{#1}[1][]{%
    \begingroup
    \colorlet{ActiveStatementAccent}{#4}%
    \def\StatementBadgeKind{#2}%
    \def\StatementBadgeNumber{#3}%
    \ifstrempty{#3}{\StatementNumberedfalse}{\StatementNumberedtrue}%
    \begin{tcolorbox}[statementbox,title={##1}]%
    \AssignSocketPlug{hyp/raisedlink}{statementbox}%
    \csname statementoriginal#1\endcsname[##1]%
    \AssignSocketPlug{hyp/raisedlink}{noop}%
  }{%
    \csname endstatementoriginal#1\endcsname
    \end{tcolorbox}\endgroup
  }%
}
\wrapstatement{theorem}{Theorem}{\thetheorem}{StatementAccent}
\wrapstatement{proposition}{Proposition}{\thetheorem}{StatementAccent}
\wrapstatement{lemma}{Lemma}{\thetheorem}{StatementAccent}
\wrapstatement{introtheorem}{Main theorem}{}{StatementAccent}
\wrapstatement{definition}{Definition}{\thetheorem}{DefinitionAccent}
\wrapstatement{example}{Example}{\thetheorem}{ExplanatoryAccent}
\wrapstatement{remark}{Remark}{\thetheorem}{ExplanatoryAccent}

\newcommand{\abstractornament}{%
  \par\noindent\makebox[\linewidth][c]{%
    \begin{tikzpicture}[baseline=0pt]
      \shade[left color=white,right color=StatementAccent!60]
        (0pt,-0.25pt) rectangle (0.45\linewidth,0.25pt);
      \shade[left color=StatementAccent!60,right color=white]
        (0.45\linewidth,-0.25pt) rectangle (0.90\linewidth,0.25pt);
    \end{tikzpicture}%
  }\par
}

\newcommand{\E}{\mathbb{E}}
\newcommand{\Pp}{\mathbb{P}}
\newcommand{\R}{\mathbb{R}}
\newcommand{\ee}{\mathrm{e}}
\newcommand{\Vjoint}{V_{\mathrm{joint}}}
\newcommand{\Vprod}{V_{\mathrm{prod}}}
\newcommand{\1}{\mathbf{1}}

\newcommand{\statementrefbadge}[3]{%
  \smash{\raisebox{-0.24ex}{%
    \tikz[baseline=0pt]{%
      \begin{scope}
        \clip[rounded corners=2.3pt] (0pt,0pt) rectangle (16.4pt,11.6pt);
        \fill[white] (0pt,0pt) rectangle (16.4pt,6.0pt);
        \fill[#1!15] (0pt,6.0pt) rectangle (16.4pt,11.6pt);
      \end{scope}
      \draw[#1!42,line width=0.25pt] (0pt,6.0pt) -- (16.4pt,6.0pt);
      \draw[#1!55,line width=0.3pt,rounded corners=2.3pt] (0pt,0pt) rectangle (16.4pt,11.6pt);
      \node[inner sep=0pt,outer sep=0pt,anchor=south west] at (0pt,0pt) {%
        \vbox to 11.6pt{\offinterlineskip
          \hbox to 16.4pt{\vbox to 5.6pt{\vfil\hbox to 16.4pt{\hfil\textcolor{#1}{\statementheadingfont\fontsize{4.9}{4.9}\selectfont #2}\hfil}\vfil}}%
          \hbox to 16.4pt{\vbox to 6.0pt{\vfil\hbox to 16.4pt{\hfil\textcolor{#1}{\statementheadingfont\fontsize{6.1}{6.1}\selectfont #3}\hfil}\vfil}}%
        }%
      };%
    }%
  }}%
}
\newrobustcmd{\namedstatementref}[4][]{%
  \hyperref[#4]{\textcolor{#2}{\ifstrempty{#1}{\nameref*{#4}}{#1}}\nobreak\hspace{1.6pt}\statementrefbadge{#2}{#3}{\ref*{#4}}}%
}
\newrobustcmd{\thmref}[2][]{\namedstatementref[#1]{StatementAccent}{Thm}{#2}}
\newrobustcmd{\propref}[2][]{\namedstatementref[#1]{StatementAccent}{Prop}{#2}}
\newrobustcmd{\lemref}[2][]{\namedstatementref[#1]{StatementAccent}{Lem}{#2}}
\newrobustcmd{\defref}[2][]{\namedstatementref[#1]{DefinitionAccent}{Def}{#2}}

\allowdisplaybreaks
\titlespacing*{\section}{0pt}{2.4ex plus 0.6ex minus 0.3ex}{1.0ex}
\titlespacing*{\subsection}{0pt}{1.8ex plus 0.5ex minus 0.2ex}{0.7ex}
\setlist[itemize]{leftmargin=2em,itemsep=0.25em,topsep=0.35em}
\setlist[enumerate]{leftmargin=2.2em,itemsep=0.25em,topsep=0.35em}

\newcommand{\authorblock}[3]{%
  \begin{tabular}[t]{@{}c@{}}
    {\fontsize{12}{14.5}\selectfont #1}\\[3pt]
    {\fontsize{8.5}{11}\selectfont\color{ContentsHeading} #2}\\[1.5pt]
    {\fontsize{8.5}{11}\selectfont\href{mailto:#3}{\nolinkurl{#3}}}
  \end{tabular}%
}

\makeatletter
\newif\if@tocaftersub
\newif\if@tocfirst
\patchcmd{\@maketitle}{\null}{\vspace*{-2.2em}}{}{}
\patchcmd{\@maketitle}{\LARGE \@title}{\LARGE\bfseries \@title}{}{}
\newcommand{\twocolumntoc}{%
  \begingroup
  \fontsize{8.5}{11.2}\selectfont
  \setlength{\parskip}{0pt}%
  \color{ContentsGray}%
  \hypersetup{linkcolor=ContentsGray}%
  \global\@tocaftersubfalse
  \global\@tocfirsttrue
  \renewcommand{\@dotsep}{4.5}%
  \renewcommand{\@pnumwidth}{1.35em}%
  \renewcommand*\l@section[2]{%
    \ifnum \c@tocdepth >\z@
      \addpenalty\@secpenalty
      \if@tocfirst
        \global\@tocfirstfalse
      \else
        \if@tocaftersub
          \addvspace{0.72em}%
        \else
          \addvspace{0.21em}%
        \fi
      \fi
      \global\@tocaftersubfalse
      \setlength\@tempdima{1.4em}%
      \begingroup
        \parindent \z@ \rightskip \@pnumwidth
        \parfillskip -\@pnumwidth
        \leavevmode
        \advance\leftskip\@tempdima
        \hskip -\leftskip
        {\bfseries\color{ContentsHeading}\hypersetup{linkcolor=ContentsHeading}##1}%
        \nobreak\hspace{0.35em}\leaders\hbox{$\m@th\mkern \@dotsep mu\hbox{\textcolor{ContentsGray!30}{.}}\mkern \@dotsep mu$}\hfill
        \nobreak\hb@xt@\@pnumwidth{\hss\bfseries\color{ContentsHeading}\hypersetup{linkcolor=ContentsHeading}##2}\par
      \endgroup
    \fi}%
  \renewcommand*\l@subsection[2]{%
    \addvspace{0.04em}%
    \global\@tocaftersubtrue
    \global\@tocfirstfalse
    \@dottedtocline{2}{1.4em}{1.8em}{##1}{##2}}%
  \patchcmd{\@dottedtocline}{\leaders}{\hspace{0.35em}\leaders}{}{}%
  \patchcmd{\@dottedtocline}{\hbox{.}}{\hbox{\textcolor{ContentsGray!36}{.}}}{}{}%
  \appto\columnbreak{\global\@minipagetrue\global\@tocaftersubfalse\global\@tocfirsttrue}%
  \setlength{\multicolsep}{0pt}%
  \setlength{\columnsep}{26pt}%
  \setlength{\columnseprule}{0.35pt}%
  \renewcommand{\columnseprulecolor}{\color{StatementAccent!18}}%
  \noindent\makebox[\linewidth][c]{%
    \begin{minipage}{0.90\linewidth}%
      \begin{multicols}{2}%
        \@starttoc{toc}%
      \end{multicols}%
    \end{minipage}%
  }\par
  \endgroup
}
\makeatother

\newcommand{\paperfrontmatter}[1]{%
  \hypersetup{pageanchor=false}%
  \pagenumbering{gobble}%
  \maketitle
  \thispagestyle{empty}%
  \vspace{-10pt}%
  \abstractornament
  \vspace{4pt}%
  #1
  \vspace{1pt}%
  \abstractornament
  \vspace{16pt}%
  \twocolumntoc
  \clearpage
  \pagenumbering{arabic}%
  \hypersetup{pageanchor=true}%
}

\newcommand{\unnumberedsection}[1]{%
  \section*{\phantomsection #1}%
  \addcontentsline{toc}{section}{\protect\numberline{}#1}%
}

\newcommand{\paperbibliography}[2]{%
  \begingroup
  \small
  \patchcmd{\thebibliography}{\section*{\refname}}{\unnumberedsection{\refname}}{}{}%
  \bibliographystyle{#1}%
  \bibliography{#2}%
  \endgroup
}